\documentclass[11pt,a4paper]{article} % 10pt is ignored!
\pdfoutput=1
\usepackage{multicol}
\usepackage{jheppub}
\usepackage{amssymb}
\usepackage{graphicx}
\usepackage{amsmath}
\newcommand{\ba}{\begin{eqnarray}}
\newcommand{\ea}{\end{eqnarray}}
\newcommand{\be}{\begin{equation}}
\newcommand{\ee}{\end{equation}}
\renewcommand{\log}{\ln}
\makeatletter
\def\fmslash{\@ifnextchar[{\fmsl@sh}{\fmsl@sh[0mu]}}
\def\fmsl@sh[#1]#2{%
  \mathchoice
    {\@fmsl@sh\displaystyle{#1}{#2}}%
    {\@fmsl@sh\textstyle{#1}{#2}}%
    {\@fmsl@sh\scriptstyle{#1}{#2}}%
    {\@fmsl@sh\scriptscriptstyle{#1}{#2}}}
\def\@fmsl@sh#1#2#3{\m@th\ooalign{$\hfil#1\mkern#2/\hfil$\crcr$#1#3$}}
\makeatother
\newcommand{\zS}[0]{\fmslash{z\;\!}\:\!}

\title{
Form factors of  $B, B_{s}\to \eta^{(\prime)}$ and $D, D_{s}\to \eta^{(\prime)}$ transitions from QCD light-cone sum rules
}

\author[]{G.~Duplan\v ci\'c,}
\author[1]{B.~Meli\'c\note{Corresponding author.}}

\affiliation[]{Rudjer Boskovic Institute, Division of Theoretical Physics, \\Bijeni\v cka 54, HR-10000 Zagreb, Croatia}

\emailAdd{gorand@irb.hr}
\emailAdd{melic@irb.hr}

\abstract{
In the framework of the QCD light-cone sum rules (LCSRs) we present the analysis of all $B, B_{s}\to \eta^{(\prime)}$ and $D, D_{s}\to \eta^{(\prime)}$ form factors ($f^+, f^0$ and $f^T$) by including $m_{\eta^{(\prime)}}^2$ corrections in the leading (up to the twist-four) and next-to-leading order (up to the twist-three) in QCD, and two-gluon contributions to the form factors at the leading twist. The SU(3)-flavour breaking corrections and the axial anomaly contributions to the distribution amplitudes are also consistently taken into account. The complete results for the $f^0$ and $f^T$ form factors of $B,B_s \to \eta^{(\prime)}$ and $D, D_{s} \to \eta^{(\prime)}$ relevant for processes like $B \to \eta^{(\prime)} \tau \nu_{\tau}$ or $B_{s} \to \eta^{(\prime)} l^+ l^-$  are given for the first time, as well as the two-gluon contribution to the tensor form factors. 
The values obtained for the $f^+$ form factors are as follows:
$f^+_{B\eta}(0)= 0.168^{+0.042}_{-0.047}$, $|f^+_{B_s\eta}(0)|= 0.212^{+0.015}_{-0.013}$, $f^+_{B\eta^\prime}(0)= 0.130^{+0.036}_{-0.032}$, $f^+_{B_s\eta^\prime}(0)= 0.252^{+0.023}_{-0.020}$ and $f^+_{D\eta}(0)= 0.429^{+0.165}_{-0.141}$, $|f^+_{D_s\eta}(0)|= 0.495^{+0.030}_{-0.029}$, $f^+_{D\eta^\prime}(0)= 0.292^{+0.113}_{-0.104}$, $f^+_{D_s\eta^\prime}(0)= 0.558^{+0.047}_{-0.045}$. Also phenomenological predictions for semileptonic  $B, B_{s}\to \eta^{(\prime)}$ and $D, D_{s}\to \eta^{(\prime)}$ decay modes are given.  
}

\begin{document}
\maketitle
\flushbottom

%%%%%%%%%%%%%%%%%%%%%%%%%%
\section{Introduction} 
%%%%%%%%%%%%%%%%%%%%%%%%%%

In the view of the numerous precise new measurements of two-body nonleptonic and semileptonic $B, B_s$ and $D, D_s$ decays to $\eta^{(\prime)}$ performed by BaBar and Belle recently \cite{PDG2014} and the upcoming experimental precision in the next-generation experiments it is timely to provide precise predictions for $B, B_{s}\to \eta^{(\prime)}$ and $D, D_{s}\to \eta^{(\prime)}$ form factors for analysis of these decays. The form factors parametrize hadronic matrix elements of quark currents and describe the long-distance QCD effects in semi-leptonic and non-leptonic decays. 

All those decays are important for testing and understanding the Standard Model flavour interactions, in particular for our understanding of the QCD dynamics in the flavour physics as well as the flavour mixing given by the Cabibbo-Kobayashi-Maskawa (CKM) mixing matrix. The $B,B_s$ and $D,D_s$ decays to $\eta,\eta^{\prime}$ pseudoscalar mesons can be used to shed some light on both of these phenomena.  

Specially, the decays $B, B_s \to X_{c\bar{c}} P$, where $X_{c\bar{c}} = J/\psi, \psi, \eta^{\prime}_c, \chi_{c0,c1,c2}, h_c$ and 
$P$ is the light pseudoscalar meson $P= \pi, K, \eta, \eta^{\prime}$ are important for our understanding of the factorization hypothesis and of the origin of the nonfactorizable contributions. Namely, there is a huge discrepancy between the experimental results for some of the decays and the theoretical predictions based on the factorization. Even the inclusion of calculated nonfactorizable contributions in some of  $B \to X_{c\bar{c}} K$ decays \cite{MelicCHARM} has not shown satisfactory agreement with the experiment. Recently we have extracted the decay constants of charmonia states by LCSR and by the lattice calculations \cite{BecirevicJa}. With the determined form factors of transitions $B,B_s \to \eta^{(\prime)}$ in this paper it will be possible to analyze consistently nonleptonic decays to charmonia and to test the factorization hypothesis in such transitions.     

Decays $B_s \to X_{c\bar{c}} P$ are also useful to access CP violation in the $B_s$ sector and the phase of the $B_s-\bar{B}_s$ mixing, 
$\beta_s = {\rm arg}\left (-V_{ts}V^{\ast}_tb / V_{cs} V_{cb}^{\ast} \right )$ \cite{ColangeloFazioWang} and in the combination with the 
$B \to X_{c\bar{c}} \eta^{(\prime)}$ observables they can be also used for the determination of the $\eta-\eta^{\prime}$ mixing 
parameters \cite{FleischerKnegjensRicciardi,expJpsi}.  

By using the huge amount of data it could be possible to make a thorough analysis and to extract the nonfactorizable contributions 
of nonleptonic decays from the data. The first ingredient for the analysis is certainly our knowledge of the $B (D) \to P$ and $B_s (D_s) \to P$ form factors. These form factors have been calculated for years by using the QCD light-cone sum rule (LCSR) method \cite{Braun_old} and on the lattice, step by step improving the precision of the results. The form factors for $B (D) \to \pi,K$ and $B_s (D_s) \to \pi,K$ are known now with quite a remarkable precision due to the consistent inclusion of corrections up to the twist-four at the LO and up to the twist-3 at the NLO \cite{BZ1,AK_D,DKMMO,DupliJa}. 

With the recent update on the $\eta, \eta^{\prime}$ DAs where the SU(3) breaking effects are included consistently to the power-suppressed twist-four corrections \cite{Braun2014}, it is possible now to analyze  
$B(D) \to \eta^{(\prime)}$ and $B_s (D_s) \to \eta^{(\prime)}$  form factors to the same precision as for the  $B (D) \to \pi,K$ and $B_s (D_s) \to \pi,K$. 
But, $\eta$ and $\eta^{\prime}$ mesons exhibit some issues which makes them quite different form the pion. In the exact SU(3) flavor limit $\eta$ is a pure flavor-octet state, while $\eta^{\prime}$ is a pure flavor-singlet. Due to the 
existence of the axial $U(1)$ anomaly, i.e. the SU(3) breaking effects which are large and responsible for the heaviness of $\eta^{\prime}$, there is a mixture between flavour-octet and flavour-singlet states usually described by the mixing matrix. In addition, the flavour-singlet states can mix with the two-gluon states producing the large gluonic admixture in $\eta^{\prime}$ mesons (which are primarily flavour-singlet states) and almost negligible ones in $\eta$ mesons. 
These gluonic contributions to the $B(D) \to \eta^{(\prime)}$ and $B_s (D_s) \to \eta^{(\prime)}$  form factors enter at 
the NLO level which makes them quite nontrivial for calculation. The only existing calculation was done by  Ball and Jones \cite{BallJones} for the $f^+$ form factor of the $B \to \eta^{(\prime)}$ decay.  

We check those results, improve them by including the $m_{\eta^{(\prime)}}$ corrections to the both, the hard scattering amplitude and to the DA of $\eta^{(\prime)}$ and consistently combine them inside the $\eta-\eta^{\prime}$ mixing schemes with the 'standard' quark contributions to predict $B \to \eta^{(\prime)}$ but also $D \to \eta^{(\prime)}$ and $B_s (D_s) \to \eta^{(\prime)}$ transition form factor $f^+$. 
In order to calculate consistently rare semileptonic $B(D) \to \eta^{(\prime)}$ and $B_s (D_s) \to \eta^{(\prime)}$ decays such as, for example, $B_s \to \eta^{(\prime)} l^+l^-$ and $B_s \to \eta^{(\prime)} \nu \bar{\nu}$,  it is necessary to calculate also other form factors, $f^0$ and $f^T$ (for definitions see (\ref{eq:fplBpi},\ref{eq:fsigBpi},\ref{eq:f0})) of these decays which is for the first time done in this paper.

%%%%%%%%%%%%%%%%%%%%%%%%%%%%%%%%%%%%%%%%%%%%%%%%%%%%%%%%%%%%%%%%%%%%%%%%%%
\section{$\eta - \eta^{\prime}$ mixing schemes and distribution amplitudes} 

\subsection{Mixing}

To analyze $\eta$ and $\eta^\prime$ states, we have to deal with several definitions of matrix elements of the flavour-diagonal axial vector and pseudoscalar current:
\begin{eqnarray}
\langle P(p) | \overline{q} \gamma^{\mu} \gamma_5 q | 0 \rangle = - \frac{i}{\sqrt{2}} f_P^q p^{\mu} , \qquad  2 m_q \langle P(p) | \overline{q} \gamma_5 q | 0 \rangle = - \frac{i}{\sqrt{2}} h_P^q  ,\nonumber \\
\langle P(p) | \overline{s} \gamma^{\mu} \gamma_5 s | 0 \rangle = - i f_P^s p^{\mu} , \qquad  2 m_s \langle P(p) | \overline{s} \gamma_5 s | 0 \rangle = - i h_P^s  ,
\end{eqnarray}   
where $q= u,d$ and the isospin limit is taken, $m_q = \frac{1}{2} (m_u + m_d)$. There is also a $U(1)_A$ anomaly, 
\begin{eqnarray}
\langle P(p) |\frac{\alpha_s}{4\pi} G_{\mu\nu}^A \tilde{G}^{A,\mu\nu} | 0 \rangle = a_P \,,
\end{eqnarray} 
which is connected with derivatives of the currents through the equation of motion as 
\begin{eqnarray}
\partial_\mu (\overline{q} \gamma^\mu \gamma_5 q ) = 2 i m_q \overline{q} \gamma_5 q - \frac{\alpha_s}{4\pi} G_{\mu\nu}^A \tilde{G}^{A,\mu\nu}
\end{eqnarray} 
and included in $h_P^{q,s}$ as 
\begin{eqnarray}
a_P = \frac{h_P^q - f_P^q m_P^2}{\sqrt{2}} = h_P^s - f_P^s m_P^2 \,.
\end{eqnarray} 
In the exact SU(3) flavour-symmetry limit $a_P = 0$. 

It is known that the SU(3) breaking corrections for $\eta$ and $\eta^\prime$ are large and that 
$\eta$ and $\eta^\prime$ mix since they are not pure a flavour-octet and a flavour-singlet states, respectively.  

The mixing of $\eta$ and $\eta^\prime$ mesons is established in two mixing schemes: the singlet-octet (SO) and the quark-flavour (QF) scheme. Each of the schemes has some advantages and some disadvantages. 

In the SO scheme the mixing occurs among $SU(3)_F$ singlet $|\eta_1 \rangle = 1/\sqrt{3} |u \bar{u} + d \bar{d} + s \bar{s} \rangle$ and octet $|\eta_8 \rangle = 1/\sqrt{6} |u \bar{u} + d \bar{d} -2 s \bar{s} \rangle$ components. 
By defining the coupling of the axial-currents to $\eta$ and  $\eta^\prime$ mesons as 
\be
\langle 0 | J_{5\nu}^{i} |\eta^{(\prime)}(p) \rangle = i f_{\eta^{(\prime)}}^i p_{\mu} \,, \quad (i = 1,8)\,,
\ee
the decay constants of pure (hypothetical) singlet and octet states $f_i$ are related to the $f_{\eta^{\prime}}^i$ via two-parameter mixing matrix
\ba
\left ( \begin{matrix}
 f_\eta^8 & f_\eta^1 \\
 f_{\eta^\prime}^8  & f_{\eta^\prime}^1 \\
 \end{matrix}
 \right )  = \left ( \begin{matrix}
 \cos \theta_8 & - \sin \theta_1 \\
 \sin \theta_8  &  \cos \theta_1 \\
 \end{matrix}
 \right ) \left ( \begin{matrix}
 f_8 & 0 \\
 0  &  f_1 \\
 \end{matrix}  \right ) \,.
\ea
Since only singlet component mixes with the gluonic contributions, the renormalization scale dependence of parameters is diagonalized in the SO scheme and therefore is suitable for the analysis of the gluon distribution amplitudes \cite{KrollPassek}. Moreover, $f_8$ is scale independent and $f_1$ renormalizes multiplicatively: 
\ba
f_P^8 (\mu) &=& f_P^8 (\mu_0) \nonumber \\
f_P^1 (\mu) &=& f_P^1 (\mu_0) \left ( 1 + \frac{2 n_f}{\pi \beta_0} \left [ \alpha_s(\mu) - \alpha_s(\mu_0) \right ] \right )\,,
\ea
where $\mu_0 = 1$ GeV, the scale at which the values of the mixing parameters are determined \cite{FeldmannKrollStech}. 

The simpler mixing scheme is QF scheme. There the basic components are 
$|\eta_q \rangle = 1/\sqrt{2} |u \bar{u} + d \bar{d}\rangle$ and $|\eta_s \rangle =  |s \bar{s}\rangle$ states and the decay constants are defined as 
\be
\langle 0 | J_{5\nu}^{r} |\eta^{(\prime)}(p) \rangle = i f_{\eta^{(\prime )}}^r p_{\mu} \,, \quad (r = q,s) \,.
\ee   
Their mixing with the decay constants of pure (hypothetical) non-strange and strange states, $f_q$ and $f_s$ respectively, is given by
\ba
\left ( \begin{matrix}
 f_\eta^q & f_\eta^s \\
 f_{\eta^\prime}^q  & f_{\eta^\prime}^s \\
 \end{matrix}
 \right )  = \left ( \begin{matrix}
 \cos \theta_q & - \sin \theta_s \\
 \sin \theta_q  &  \cos \theta_s \\
 \end{matrix}
 \right ) \left ( \begin{matrix}
 f_q & 0 \\
 0  &  f_s \\
 \end{matrix}  \right ) \,.
\ea
The main advantage of this scheme is that the mixing is not governed by the (large, $10-20\%$) $SU(3)_F$ breaking effects as in the SO scheme, but by the OZI-rule violating contributions which have be proven to be small \cite{FeldmannKrollStech}. Therefore it is possible to parametrize the mixing just with one angle $\phi$ and the matrix $U(\phi)$ given as
\ba
 U(\phi) = \left ( \begin{matrix}
 \cos \phi & - \sin \phi \\
 \sin \phi  &  \cos \phi \\
 \end{matrix} \right ) 
\ea 
which leads to the following expressions
\ba
& & \left ( \begin{matrix}
 \eta\\
 \eta^\prime \\
 \end{matrix}
 \right ) = U(\phi) \left ( \begin{matrix}
 \eta_q  \\
 \eta_s \\
 \end{matrix}  \right )
 \nonumber \\
& & \left ( \begin{matrix}
 f_\eta^{(q)} &   f_\eta^{(s)} \\
  f_{\eta^\prime}^{(q)} &   f_{\eta^\prime}^{(s)} \\
 \end{matrix} \right ) = U(\phi) \left ( \begin{matrix}
 f_q & 0 \\
 0  & f_s \\
 \end{matrix}  \right )
 \, , 
 \\
&& \begin{matrix}
 f_\eta^{(q)} =  f_q \cos  \phi = \frac{1}{\sqrt{3}} \left ( \sqrt{2} f_\eta^1 + f_\eta^8 \right)\,, 
\qquad
 f_\eta^{(s)} = - f_s \sin \phi = \frac{1}{\sqrt{3}} \left (  f_{\eta}^1 - \sqrt{2} f_{\eta}^8 \right)\,, 
\nonumber\\
 f_{\eta^\prime}^{(q)} = f_q \sin \phi = \frac{1}{\sqrt{3}} \left ( \sqrt{2} f_{\eta^\prime}^1 + f_{\eta^\prime}^8 \right)\,, 
\qquad
 f_{\eta^\prime}^{(s)} = f_s \cos \phi = \frac{1}{\sqrt{3}} \left (  f_{\eta^\prime}^1 - \sqrt{2} f_{\eta^\prime}^8 \right)\,.
 \end{matrix}
 \label{eq:fetaq}
\ea   
The parameters have been determined by fits in \cite{FeldmannKrollStech} as 
\ba
f_q = (1.07 \pm 0.02) f_\pi \,, \qquad 
f_s = (1.34 \pm 0.06) f_\pi \,, \qquad 
\phi = 39.3^o \pm 1.0^o 
\,, 
\ea
and will be also used in this paper\footnote{There have been some recent discussions on the $\eta-\eta^\prime$ mixing parameters and all of them are in the range of $\phi$ above \cite{Ambrosino,KLOE-RicciardiBigi,lattice,expJpsi}.}.  
These values give for the parameters of the SO basis the following:
\ba
f_8 = (1.26 \pm 0.04) f_\pi \,, \;
f_1 = (1.17 \pm 0.03) f_\pi \,, \; 
\phi_8 = - (21.2^o \pm 1.6^o) \,, \;
\phi_1 = - (9.2^o \pm 1.7^o) \,, 
\nonumber \\
\ea
and the decay constants are connected as
 \ba
\left ( \begin{matrix}
 f_\eta^8 & f_\eta^1 \\
 f_{\eta^\prime}^8  & f_{\eta^\prime}^1 \\
 \end{matrix}
 \right )  = U(\phi)
 \left ( \begin{matrix}
 f_q & 0 \\
 0  &  f_s \\
 \end{matrix}  \right )
 \left ( \begin{matrix}
 \sqrt{\frac{1}{3}} & \sqrt{\frac{2}{3}} \\
-\sqrt{\frac{2}{3}} &  \sqrt{\frac{1}{3}} \\
 \end{matrix}  \right ) = 
 \left ( \begin{matrix}
0.1530 & 0.0243 \\
 -0.0595 &  0.1506 \\
 \end{matrix}  \right ) {\rm GeV} \,.
\ea
Due to the mixing of the flavour-singlet and gluonic components, in the QF scheme both $\eta_q$ and $\eta_s$ will get gluonic contributions and therefore 
also the physical $\eta$ and $\eta^\prime$ states.  
The flavour states in QF scheme and in the approximation above can be written as  \cite{KrollPassek}
\ba
|\eta^q \rangle = \frac{f_q}{2 \sqrt{2 N_c}} \left ( \psi_q (x,\mu_F) |q \bar{q} \rangle + \sqrt{2/3} \psi_g (x,\mu_F) | g g \rangle \right )
\\
|\eta^s \rangle = \frac{f_s}{2 \sqrt{2 N_c}} \left ( \psi_s (x,\mu_F) |s \bar{s} \rangle + \sqrt{1/3} \psi_g (x,\mu_F) | g g \rangle \right )
\ea
where $|q \bar{q} \rangle = (u\bar{u} + d\bar{d} )/\sqrt{2}$ and 
$\psi_q  = 1/3 (\psi_8 + 2 \psi_1)$ and $\psi_s = 1/3 (2 \psi_8 + \psi_1)$. 

By combing above information about the nature of $\eta$ and $\eta^{\prime}$ states one can expect that gluonic contributions $|gg\rangle$ will be larger for $\eta^\prime$
mesons, which is confirmed by the final results.
%, Eqs. (5.2-5.5), see $b_2^{P,g}$.

Until now there is no available QCD sum rule or lattice QCD calculations of $B_s$ to $\eta^{(\prime)}$ transition form factors $f_{B_s\eta^{(\prime)}}^{+,0,T}$. Since these transitions probe only the $|s \bar{s} \rangle $ content, one can use the approximation in the quark flavour scheme 
\ba
f_{B_s \eta} = - \sin \phi f_{BK}  \, , \qquad f_{B_s \eta^{\prime}} = \cos \phi f_{BK} \,,
\label{eq:SU3}
\ea
which neglects completely the gluonic contribution. The calculation presented in this paper will  check for the $SU(3)_F$ breaking effects in the above relations.

\subsection{Distribution amplitudes}

The light-cone distribution amplitudes (DA), giving the momentum fraction distribution of valence quarks of $\eta$ and $\eta^\prime$ are defined analogously to  other meson light-cone DAs, by expanding the non-local operators on the light-cone in terms of increasing twist, but paying attention to the specific flavour structure of $\eta^{(\prime)}$ mesons.

The twist 2 two-quark DAs $\phi_{2,P}^i$ of  $P = \eta^{(\prime)}$ mesons are defined as 
\ba
\langle 0| \bar{\Psi}(z) {\cal C}_i \zS \gamma_5 [z,-z] \Psi(-z) | P(p) \rangle = 
i (pz) f_P^i \int_0^1 du e^{i (2u -1) (pz)} \phi_{2,P}^i(u) \,,
\ea  
where as usual $z_\mu$ is the light-like vector and $[z,-z]$ is the path-ordered gauge connection and $u $ is the momentum fraction of a valence quark. 
In the SO basis one will have ${\cal C}_1 = 1/\sqrt{3}$ and ${\cal C}_8 = \lambda_8/\sqrt{2}$ ( $\lambda_8$ is the standard Gell-Mann matrix ), while in QF basis the constants are ${\cal C}_q = (\sqrt{2} {\cal C}_1 
+ {\cal C}_8 )/\sqrt{3}$ and ${\cal C}_s = ({\cal C}_1 
- \sqrt{2} {\cal C}_8 )/\sqrt{3}$.  
The twist-2 two-quark DAs of $\eta^{(\prime)}$ are symmetric in their argument and therefore they can be expanded in terms of Gegenbauer polynomials as usually:
\ba
\phi_{2,P}^i(u) = 6 u (1-u) \left ( 1 + \sum_{n=2,4,..} a_n^{P,i}(\mu) C_n^{3/2}(2u -1) \right )\, \quad (i = 1,8,q,s)\,.
\ea 
The coefficients $a_n^{P,i}$ are the Gegenbauer moments of the quark DA. 

The gluonic twist-2 DA $\phi_{2,P}^g$  of $P = \eta^{(\prime)}$ mesons are defined by the following matrix element
(for detailed discussion on the derivation of gluonic DA and its mixing with the quark states in mesons see for example \cite{baier}):
\ba
n_{\mu} n_{\nu} \langle 0| G^{\mu\alpha}(z) [z,-z] \tilde{G}_{\alpha}^{\nu}(-z) | P(p) \rangle = 
\frac{1}{2} \frac{C_F}{\sqrt{3}} (pz)^2 f_P^i \int_0^1 du e^{i (2u -1) (pz)} \phi_{2,P}^g(u)\,.
\ea
It is antisymmetric and therefore 
\ba
\phi_{2,P}^g(u) =  - \phi_{2,P}^g(1-u) \,,
\ea
and it is expanded in terms of $C_n^{5/2}$ Gegenbauer polynomials
\ba
\phi_{2,P}^g(u) = u^2 (1-u)^2 \left ( \sum_{n=2,4,..} b_n^{P,g}(\mu) C_{n-1}^{5/2}(2u -1) \right ) \,,
\ea
where the coefficients $b_n^{P,g}$ are the Gegenbauer moments of the gluon DA and we take $b_n^{\eta,g} = b_n^{\eta^{\prime},g}$
and keep only the first term in the sum, $n=2$. Although $b_n^{\eta,g}$ and $b_n^{\eta^{\prime},g}$ could differ, this approximation is justified 
since their values are subject of large uncertainties.

In the calculation we use the following matrix element of the $\eta^{(\prime)}$ over two gluon fields 
\begin{equation}
\langle 0| A^A_{\alpha}(z) A^B_{\beta}(-z)|P(p)\rangle = \frac{1}{4} \epsilon_{\alpha \beta \rho \sigma}
\frac{z^{\rho}p^{\sigma}}{(pz)}\frac{C_F}{\sqrt{3}}f^1_P\frac{\delta^{AB}}{8}\int_0^1 du\,e^{i\xi(pz)}\frac{\phi^g_{2,P}(u)}{u(1-u)}\,.
\label{dodform}
\end{equation}

With the above normalization of the DA, the renormalization mixing of twist-2 quark and gluonic distribution amplitudes is given as 
\ba
\mu \frac{d}{d \mu} 
\left ( \begin{matrix} a_2^{\eta^{(\prime)},1} \\ b_2^{\eta^{(\prime)} ,g}\\ \end{matrix} \right ) 
= -\frac{\alpha_s(\mu)}{4 \pi} \left ( \begin{matrix}
 \frac{100}{9} & - \frac{10}{81}  \\
 -36   &  22 \\
 \end{matrix}
 \right ) \left ( \begin{matrix} a_2^{\eta^{(\prime)},1} \\ b_2^{\eta^{(\prime)} ,g} \\ \end{matrix} \right )
\label{eq:evol}
\ea
and it is numerically small. But, the mixing is important for $p^2 =0$ case, since it verifies the collinear 'factorization formula'  for the form factors
\ba
F(q^2, (p+q)^2)  = \int_0^1 du\,\sum_n T_H^{(n)} (u, q^2, (p+q)^2,\mu_{\rm IR}) \psi_{n}(u,\mu_{\rm IR})\,, 
\ea
and proves that the separation of the transition form factors in perturbatively calculable hard-scattering $T_H$ part and a nonperturbative DA is essentially independent on the factorization scale $\mu_{\rm IR}$ \cite{MelicMALI}. 
This is an essential step of calculation which is going to be proved for each of the $F$-correlation function at the order of twist 2, see discussion in the next section.

The explicit solutions of (\ref{eq:evol}) can be find in \cite{BallJones} and in the appendix B of \cite{Braun2014}. 

In the asymptotic case, when $Q^2 = -q^2 \to \infty$ the twist-2 quark and gluon DAs evolve to their asymptotic forms 
\ba
\phi_{2,P}^i(u)_{|{\rm asym}} &=& 6 u (1-u)\,,
\nonumber  \\
\phi_{2,P}^g(u)_{|{\rm asym}} &=& 0 \,.
\ea
In that case, there is no gluonic contribution at the twist-2 level to the form factors, and the residual $\mu_{\rm IR}$ dependence in the twist-2 NLO quark contribution integrates with $\phi_{2,P}^i(u)_{|{\rm asym}}$ to zero, which again confirms the $\mu_{IR}$ independence of the complete result.  

To include SU(3) flavour-breaking corrections consistently we keep not only $m_{\eta^{(\prime)}}^2$ corrections and quark masses in the hard-scattering amplitudes, but also in the distribution amplitudes.   Therefore we do not use the approximations in the twist-3 and twist-4 contributions employed in the literature where the following replacements are used in DAs: 
\begin{eqnarray}
 f_{\pi} \frac{m_{\pi}^2}{2 m_q} \to f_{q} \frac{m_{\pi}^2}{2 m_q} \,, \qquad 
  f_{\pi} \frac{m_{\pi}^2}{2 m_s} \to f_{s} \frac{2m_{K}^2 - m_{\pi}^2}{2 m_s}  \,
\label{eq:hq-approx} 
\end{eqnarray} 
for $M \to \eta_q$ and $M \to \eta_s$ decays respectively. Instead we are going to use (in the QF scheme): 
\begin{eqnarray}
&& f_{\pi} m_{\pi}^2 \to h_{q} = f_q (m_\eta^2 \cos^2 \phi + m_{\eta^{\prime}}^2 \sin^2 \phi) - \sqrt{2} f_s (m_{\eta^{\prime}}^2 - m_\eta^2) \sin \phi \cos \phi \, ,
 \nonumber \\
&& f_{\pi} m_{\pi}^2 \to h_{s} = f_s (m_{\eta^{\prime}}^2 \cos^2 \phi + m_{\eta}^2 \sin^2 \phi) - \frac{f_q}{\sqrt{2}} (m_{\eta^{\prime}}^2 - m_\eta^2) \sin \phi \cos \phi  \, .  
\end{eqnarray} 
Although the above quantities, especially $h_q$, are weakly constrained due to the numerical cancellations, 
\ba
h_{q} = 0.0015 \pm 0.0040 \,{\rm GeV}^3, \quad h_{s} = 0.087 \pm 0.006 \,{\rm GeV}^3, 
\label{eq:hqhs}
\ea
we use them for the consistency of our calculation. Actually, we will see later that the approximation in (\ref{eq:hq-approx}) for $h_q$ is quite bad and causes somewhat  large values of form factors of $D, B \to \eta^{(\prime)}$. 

Distribution amplitudes of higher twist are defined following \cite{Braun2014} and \cite{BBL}. 
Their parameter evolutions and definitions include now the anomaly contribution $a_P$ with the following expressions \cite{BenekeNeubert}: 
\ba
a_\eta = - \frac{1}{\sqrt{2}} ( f_q m_\eta^2 - h_q) \cos\phi = - \frac{ m_{\eta^{\prime}}^2 -m_\eta^2}{\sqrt{2}} \sin\phi \cos\phi \left ( - f_q \sin\phi + \sqrt{2} f_s \cos\phi \right) \,, \nonumber \\
a_{\eta^{\prime}} = - \frac{1}{\sqrt{2}} ( f_q m_{\eta^{\prime}}^2 - h_q) \sin\phi = - \frac{ m_{\eta^{\prime}}^2 -m_\eta^2}{\sqrt{2}} \sin\phi \cos\phi \left (f_q \cos\phi + \sqrt{2} f_s \sin\phi \right)\,. \nonumber \\
\ea
Therefore in \cite{Braun2014} the normalizations of two-particle twist-3 DAs $\phi_{3}^{p,\sigma}$  differ from those in \cite{BBL}. In \cite{Braun2014} one can find a consistent treatment of $m_s$ corrections up to twist-4 and of anomalous contributions to DA and we take definitions and expressions given there. 
%Therefore we will repeat some definitions here. More detailed discussion can be found in %\cite{Braun2014} and \cite{BBL}. 

Then, 
\ba
2 m_r \langle 0 | \bar{r}(z_2 n) i \gamma_5 r (z_1n) |P(p) \rangle = \int_0^1 du  e^{-i (u z_1 + \bar{u} z_2)(pn) } \phi_{3P}^{(r)p} (u) \,,
\ea
where $r = q,s$ and 
\ba
2 m_r \langle 0 | \bar{r}(z_2 n) \sigma_{\mu\nu}\gamma_5  r (z_1 n) |P(p) \rangle = \frac{i (z_1 - z_2)}{6} (p_{\mu} n_{\nu} - p_{\nu} n_{\mu} ) \int_0^1 e^{-i (u z_1 + \bar{u} z_2)(pn)} \phi_{3P}^{(r) \sigma} (u) \,. 
\nonumber \\
\ea
The normalization is then 
\ba
\int_0^1 du \phi_{3P}^{(r)p} (u) = \int_0^1 du \phi_{3P}^{(r) \sigma} (u) = H_P^{(r)} \,,
\ea
where 
\begin{equation}
H_P^{(r)} = m_P^2 F^{(r)}_P - a_P \,, \qquad  H_P^{(q)} = \frac{h_P^{(q)}}{\sqrt{2}}\, , H_P^{(s)} = h_P^{(s)}\, , 
\end{equation}
and 
\begin{equation}
F_P^{(q)} = \frac{f_P^{(q)}}{\sqrt{2}}\,, \quad  F_P^{(s)} = f_P^{(s)}\, .
\end{equation}

By calculating the mixing of twist-4 DAs, some approximations in the twist-3 DA are made in \cite{Braun2014} when compared to the expressions in \cite{BBL}, 
to keep the same order of calculation in the conformal spin and the quark masses. 

The expressions for the two-particle twist-3 DAs used 
(contributions of higher conformal spin and $O(m_s^2)$ corrections are neglected; see also \cite{BBL}, Eqs.(3.25-26))
are 
\ba
\phi_{3s}^p = h_s + 60 m_s f_{3s} C_2^{1/2} (2u -1) \,, \nonumber \\
\phi_{3s}^\sigma = 6 u (1-u) \left [ h_s + 10 m_s f_{3s} C_2^{3/2} ( 2 u -1) \right ] \,.
\ea 

The three-particle quark-gluon-antiquark DA is defined as usual \cite{BBL}
\ba
\langle 0 | \bar{r}(z) \sigma_{\mu\nu}\gamma_5 g G_{\alpha \beta} (vz) r(-z) |P(p) \rangle = i f_{3 r} \left (p_\alpha p_\mu g_{\nu \beta}^{\perp} - p_\alpha p_\nu g_{\mu\beta}^{\perp} - ( \alpha \leftrightarrow  \beta ) \right ) \nonumber \\
\int_0^1 d \alpha_1 d \alpha_2
d \alpha_3 \delta(1 - \alpha_1 - \alpha_2 -\alpha_3) \Phi_{3r} (\alpha_1,\alpha_2,\alpha_3) \,, 
\ea
\ba
\Phi_{3r}(\alpha) = 360 \alpha_1 \alpha_2 \alpha_3^2 \left \{ 1 + \lambda_{3r} (\alpha_1 - \alpha_2 ) + \omega_{3r} \frac{1}{2} ( 7 \alpha_3 -3 ) \right \} \,.
\ea

There are two two-particle twist-4 DAs $\psi_{4P}^{(r)}(u)$, $\phi_{4P}^{(r)}(u)$ and four three-particle twist-4 DAs, $\Psi_{4P}^{(r)}(\alpha)$, $\tilde{\Psi}^{(r)}_{4 P}(\alpha)$,  $\Phi_{4P}^{(r)}(\alpha)$, $\tilde{\Phi}_{4 P}^{(r)}(\alpha)$. All details and subtleties in derivation of these improved twist-4 DAs with the corrected mass corrections and inclusion of the anomalous contribution can be found in Appendix A of \cite{Braun2014}. Here we just quote the expressions:
\ba
\psi_{4P}^{(r)}(u) &=& \psi_{4P}^{(r){\rm tw}}(u) + m_P^2 \psi_{4P}^{(r){\rm mass}}(u)
\nonumber \\
\hspace*{0.5cm}\psi_{4P}^{(r){\rm tw}}(u) &=& \frac{20}{3} \delta_P^{2 (r)} C_2^{1/2}(2 u-1) 
+ 30 m_r \frac{f_{3P}^{(r)}}{f_P^{(r)}} \left ( \frac{1}{2} - 10 u (1-u) + 35 u^2 (1-u)^2 
\right )\,,
\nonumber \\
\hspace*{0.5cm}\psi_{4P}^{(r){\rm mass}}(u) &=& \frac{17}{12} - 19 u (1-u) + \frac{105}{2} u^2 (1-u)^2 + 
c_{2P}^{(r)} \left ( \frac{3}{2} - 54 u (1-u) + 225 u^2 (1-u)^2   \right )  \,,
\nonumber \\
\phi_{4P}^{(r)}(u) &=& \phi_{4P}^{(r){\rm tw}}(u) + m_P^2 \phi_{4P}^{(r){\rm mass}}(u) \,,
\nonumber \\
\hspace*{0.5cm}\phi_{4P}^{(r){\rm tw}}(u) &=& \frac{200}{3} \delta_P^2{(r)} u^2 (1-u)^2 + 21 \delta_P^{2(r)} 
\omega_{4P}^{(r)} \left [ u (1-u) (2 + 3 u (1-u) ) \right . 
\nonumber \\
&& \left .  + 2 ( u^3 (10 - 15 u + 6 u^2) \ln u + 
(u  \leftrightarrow (1-u) ) \right ]
\nonumber \\
&& + 20 m_r \frac{f_{3P}^{(r)}}{f_P^{(r)}} u (1-u) \left ( 12 - 63 u (1-u) + 14 u^2 (1-u)^2 
\right )\,,
\nonumber \\
\hspace*{0.5cm}\phi_{4P}^{(r){\rm mass}}(u) &=& u (1-u) \left [ \frac{88}{15} + \frac{39}{5} u (1-u) + 
14 u^2 (1-u)^2 \right ] 
\nonumber \\
&& - c_{2P}^{(r)} u (1-u) \left ( \frac{24}{5} - \frac{54}{5} u (1-u) + 180 u^2 (1-u)^2 \right ) 
\nonumber \\
&& + \left ( \frac{28}{15} - \frac{24}{5} c_{2P}^{(r)} \right ) \left [ 
u^3 ( 10 - 15 u + 6 u^2 )\ln u + (u \leftrightarrow (1-u) ) \right ] \,,
\ea
and
\ba
\Phi_{4P}^{(r)}(\alpha) &=& 120 \alpha_1\alpha_2\alpha_3 \left [ (\alpha_1 - \alpha_2) \phi_{1,P}^{(r)} \right ]\,, 
\nonumber \\
\tilde{\Phi}^{(r)}_{4 P}(\alpha) &=& 120 \alpha_1\alpha_2\alpha_3 \left [ \tilde{\phi}_{0,P}^{(r)} + ( 3 \alpha_3 -1) \tilde{\phi}_{2,P}^{(r)}\right ] \,, 
\nonumber \\
\Psi_{4P}^{(r)}(\alpha) &=& - 30 \alpha_3^2 (\alpha_1- \alpha_2 )  \left [ \psi_{0,P}^{(r)} + \alpha_3 \psi_{1,P}^{(r)} + \frac{1}{2} ( 5 \alpha_3 -3) \psi_{2,P}^{(r)}\right ]\,, 
\nonumber \\
\tilde{\Psi}^{(r)}_{4 P}(\alpha) &=& - 30 \alpha_3^2  \left [ (1 - \alpha_3) \psi_{0,P}^{(r)} + \left( \alpha_3 (1- \alpha_3) - 6 \alpha_1\alpha_2 \right ) \psi_{1,P}^{(r)}
\right . \nonumber \\
&& \left . \hspace*{1cm} + \left (\alpha_3 (1- \alpha_3)-  \frac{3}{2} (\alpha_1^2 + \alpha_2^2) \right ) \psi_{2,P}^{(r)} \right ]\,, 
\ea
where 
\ba
 \phi_{1,P}^{(r)}  &=& \frac{21}{8} \left [ \delta_P^{2(r)} \omega_{4P}^{(r)} + 
 \frac{2}{45} m_P^2 \left ( 1 - \frac{18}{7} a_{2P}^{(s)} \right ) \right ]\,,
\nonumber \\
 \tilde{\phi}_{0,P}^{(r)} &=& - \frac{1}{3} \delta_P^{2(r)}\,,
 \nonumber \\
 \tilde{\phi}_{2,P}^{(r)} &=& \frac{21}{8} \delta_P^{2(r)} \omega_{4P}^{(r)} \,,
\nonumber \\
\psi_{0,P}^{(r)} &=& - \frac{1}{3} \delta_P^{2(r)}\,,
\nonumber \\
\psi_{1,P}^{(r)} &=&  \frac{7}{4} \left [ \delta_P^{2(r)} \omega_{4P}^{(r)} + 
 \frac{1}{45} m_P^2 \left ( 1 - \frac{18}{7} a_{2P}^{(s)} \right ) 
 + 4 m_r \frac{f_{3P}^{(r)}}{f_P^{(r)}}  \right ]\,,
\nonumber \\
\psi_{2,P}^{(r)}&=& \frac{7}{4} \left [ 2 \delta_P^{2(r)} \omega_{4P}^{(r)} -
 \frac{1}{45} m_P^2 \left ( 1 - \frac{18}{7} a_{2P}^{(s)} \right ) 
 - 4 m_r \frac{f_{3P}^{(r)}}{f_P^{(r)}}  \right ]\,.
\ea
The parameters which appear here are parametrization of various local matrix elements and their values are taken from \cite{BBL} and listed in Appendix B.

The above twist-4 expressions are valid for flavour-octet contributions where there is no mixing with the gluonic twist-4 DA. For the flavour-singlet case one has to take this mixing into account. In the approximation taken in \cite{Braun2014} the twist-4 DAs are given by the replacement
\ba
m_P^2 f_M^{(r)} \rightarrow  h_P^{(r)} = m_P^2 f_M^{(r)}  -a_P
\ea
everywhere at the twist-4 level where the mass $m_P^2$ occurs. As it was discussed in 
\cite{Braun2014} this substitution ensures for the given accuracy  the consistent normalization of the twist-3 and twist-4 DA and ensures that the same mixing FKS scheme applies also for the higher-twist contributions. 

For the values of parameters involved we will use crude estimates in terms of the pion 
and kaon DA parameters derived from the sum rules \cite{BBL}, see Appendix:
\ba
a_{2,4}^{P,q} \simeq a_{2,4}^{P,s} = a_{2,4}^{\pi} \,, \nonumber \\
f_{3q} \simeq f_{3\pi} \,, \qquad
f_{3s} \simeq f_{3K} \,,\nonumber \\
\lambda_{3q} \simeq 0 \,, \qquad
\lambda_{3s} \simeq \lambda_{3K}\,, \nonumber \\
\omega_{3q} \simeq \omega_{3\pi}\,, \qquad
\omega_{3s}\simeq \omega_{3K} \,,\nonumber \\
\kappa_{4q} \simeq 0 \,, \qquad
\kappa_{4s}\simeq \kappa_{4K} \,,\nonumber \\
\delta_{P}^{2(q)} \simeq \delta_{\pi}^2 \,, \qquad
\delta_{P}^{2(s)}\simeq \delta_{K}^2 \,,\nonumber
\ea
while the corresponding $\eta,\eta^\prime$ parameters will be given through the mixing as
\ba
\left ( \begin{matrix}
 f_{3\eta}^{(q)} &   f_{3\eta}^{(s)} \\
  f_{3\eta^\prime}^{(q)} &   f_{3\eta^\prime}^{(s)} \\
 \end{matrix} \right )  = U(\phi) \left ( \begin{matrix}
 f_{3q} & 0 \\
 0  & f_{3s} \\
 \end{matrix} \right ) \,,
 \qquad 
 \left ( \begin{matrix}
 h_{\eta}^{(q)} &  h_{\eta}^{(s)} \\
  h_{\eta^\prime}^{(q)} &   h_{\eta^\prime}^{(s)} \\
 \end{matrix} \right )  = U(\phi) \left ( \begin{matrix}
 h_{q} & 0 \\
 0  & h_{s} \\
 \end{matrix} \right ) \,.
\ea

%%%%%%%%%%%%%%%%%%%%%%%%%%%%%%%%%%%%%%%%%%%%%%%%%%%%%%%%%%%%%%%%%%%%%%%%%%
\section{LCSR for $B, B_{s}\to \eta^{(\prime)}$ and $D, D_{s}\to \eta^{(\prime)}$ form factors} 

For calculating the $M \to \eta^{(\prime)}$ form factors, where $M = B,B_s,D,D_s$, by using the LCSR method one considers a vacuum-to-$\eta$ or vacuum-to-$\eta^{\prime}$ correlation functions of a weak current and an interpolating current with the quantum numbers of a meson $M$. For $B \to \eta^{(\prime)}$,  the form  factors $f^+_{B\eta^{(\prime)}}$, $f^0_{B\eta^{(\prime)}}$ and $f^T_{B\eta^{(\prime)}}$ will be defined with the help of the correlator
\ba
F_{\mu}(p,q) &=& i \int d^4x ~e^{i q\cdot x}
\langle \eta^{(\prime)}(p)| T \left \{ \bar{u}(x)\Gamma_\mu b(x), j_B(0) \right \} |0 \rangle
\nonumber \\
&=& \Bigg \{ 
\begin{array}{ll}
F(q^2,(p+q)^2)p_\mu +\widetilde{F}(q^2,(p+q)^2)q_\mu\,,& ~~\Gamma_\mu= \gamma_\mu\\
 & \\
F^T(q^2,(p+q)^2)\big[p_\mu q^2-q_\mu (q p)\big]\,,& ~~\Gamma_\mu= -i\sigma_{\mu\nu}q^\nu\\
\end{array} 
\label{eq:corr}
\ea
for two different $b \to u$ transition currents, where $j_B = m_b\bar{b}i\gamma_5 u$. Analogous formulas are going to be valid for $D \to \eta^{(\prime)}$ with the replacements $b \to c$ and $u \to d$ in the transition currents and $j_B \to j_D = m_c\bar{c}i\gamma_5 d$. 
For $f^{+,0,T}_{B_{s}\eta^{(\prime)}}$ form factors we consider 
the replacement $u \to s$ in (\ref{eq:corr}) and $j_{B_s} = m_b\bar{b}i\gamma_5 s$ interpolating current. Again, $D_s$ case is then obtained trivially by replacing $b$-quark with the $c$-quark.  

Since we want to explore also the SU(3) symmetry breaking, we will keep the 
$\eta^{(\prime)}$ masses ($p^2 = m^2_{\eta^{(\prime)}}$) in (\ref{eq:corr}). 
The light quark masses will be systematically neglected, except when they occur in ratios in the distribution amplitudes.

The method of the LCSR is very well know  and  we will here just briefly outline the procedure in order to properly define all ingredients necessary for calculating the form factors. 
For the large virtualities of the currents above, the correlation function is dominated by the distances $x^2 =0$ near the light-cone, and 
factorizes to the convolution of the nonperturbative, universal  part (the light cone distribution amplitude (DA)) and the perturbative, short-distance part, 
the hard scattering amplitude, as a sum of contributions of increasing twist.

We calculate here contributions up to the twist-4 in the 
leading order, $O(\alpha_s^0)$, and up to the twist-3 in NLO, neglecting the three-particle contributions at this level.
Schematically, the contributions are shown in Fig.1 and Fig.2. 
%%%%%%%%%%%%%%%%%%%%%%%%%%%%%%%%%%%
%%%%%%%%%fig1: LO Diagrams
%\FIGURE[h]{
\begin{figure}[t]
\begin{center}
\includegraphics[width=3.9cm]{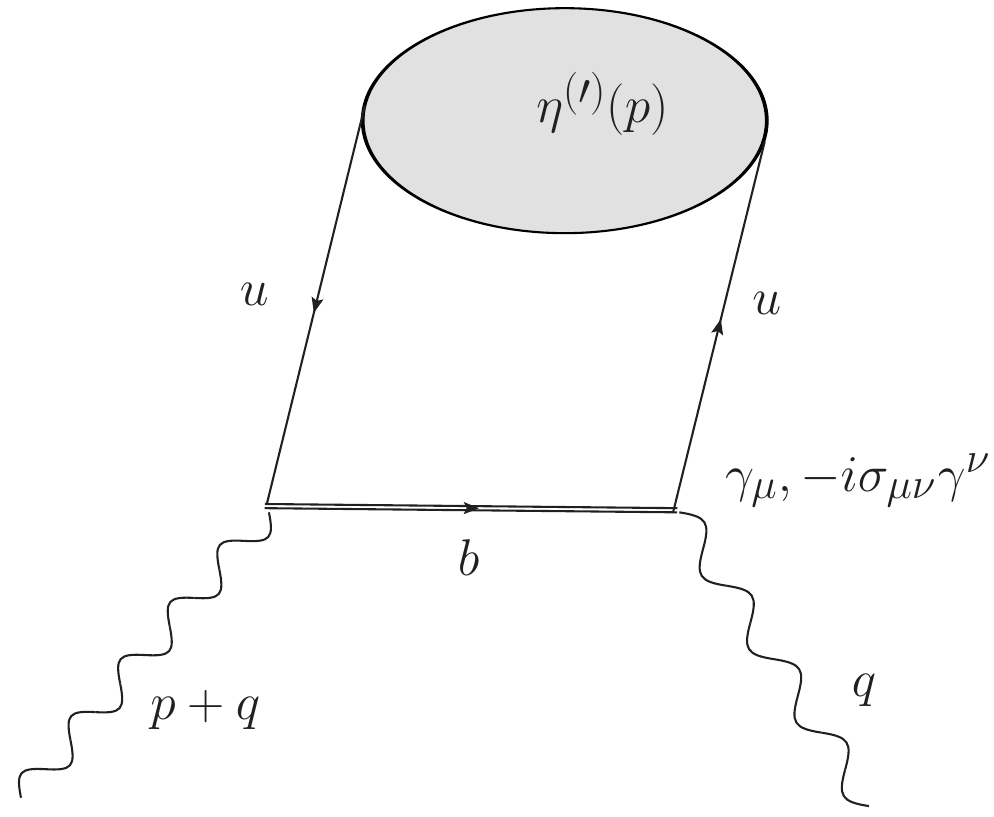}\hspace{1cm}
\includegraphics[width=3.9cm]{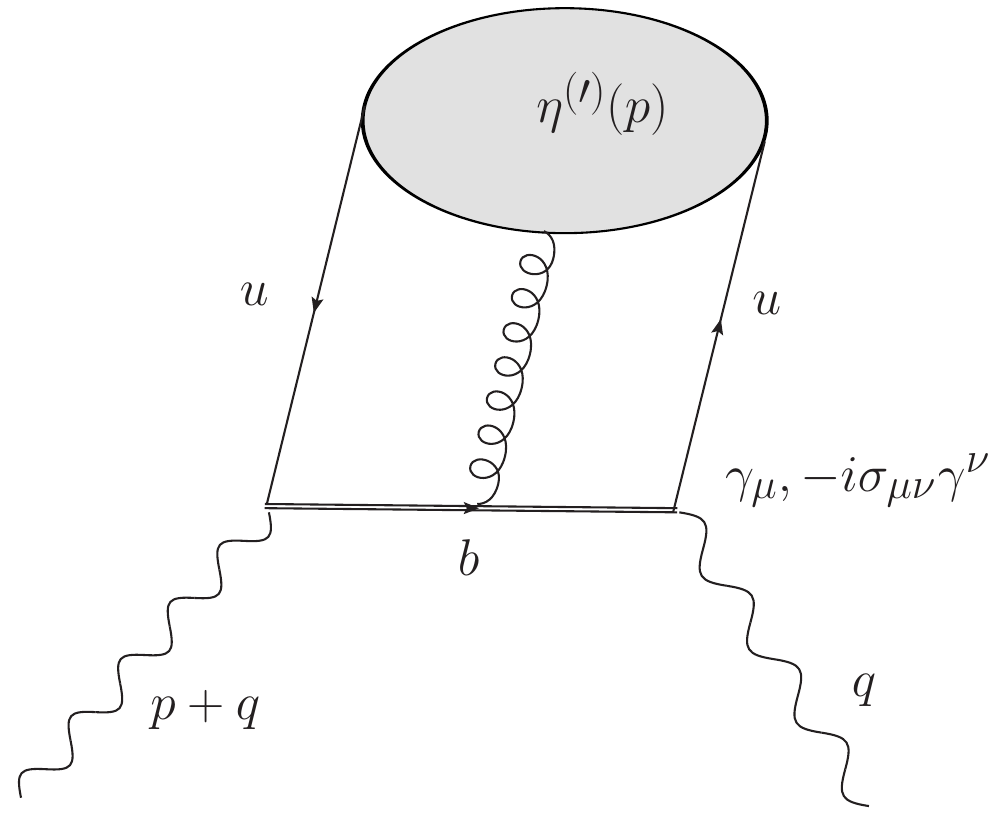}\\
\caption{ \it Diagrams corresponding to the leading-order terms in 
the hard-scattering amplitudes involving the  
two-particle (left) and three-particle (right) $\eta^{(\prime)}$ DA's shown by ovals. 
Solid, curly and wave lines represent quarks, gluons, and external 
currents, respectively. For $B_s \to \eta^{(\prime)}$ transition, $u$ is replaced by $s$. In the case of $D \to \eta^{(\prime)}$ transitions, $u \to d$ and $b \to c$ and correspondingly $d$ is exchanged by $s$ for $D_s \to \eta^{(\prime)}$. }
\label{fig-diags}
\end{center}
\end{figure}
%}
%%%%%%%%%%%%%%%%%%%%%%%%%%%%%%%%%%%
%%%%%%%%%fig2: alpha_s Diagrams
%\FIGURE[t]{
\begin{figure}[t]
\begin{center}
\includegraphics[width=12.5cm]{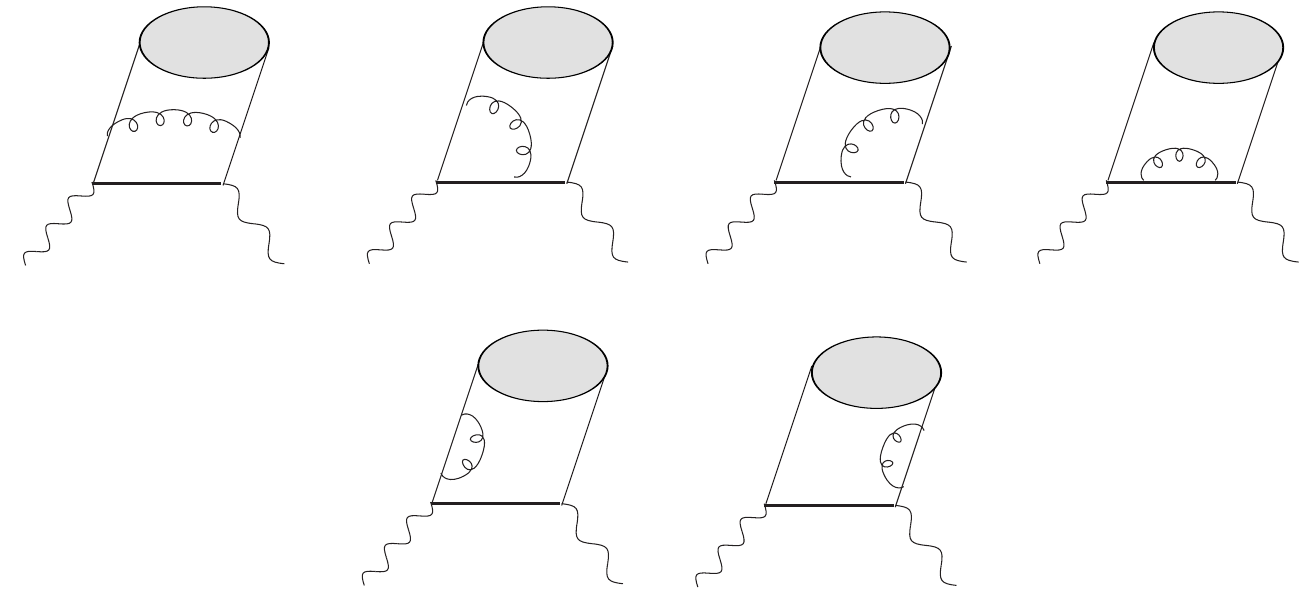}
\caption{ \it Diagrams contributing to the quark hard-scattering amplitudes at $O(\alpha_s)$.}
\label{fig-alphas}
\end{center}
\end{figure}
%}
%%%%%%%%%%%%%%%%%%%%%%%%%%%%%%%%%%%
%%%%%%%%%fig3: gluonic constributions
%\FIGURE[t]{
\begin{figure}[t]
%\begin{center}
\includegraphics[width=3.5cm]{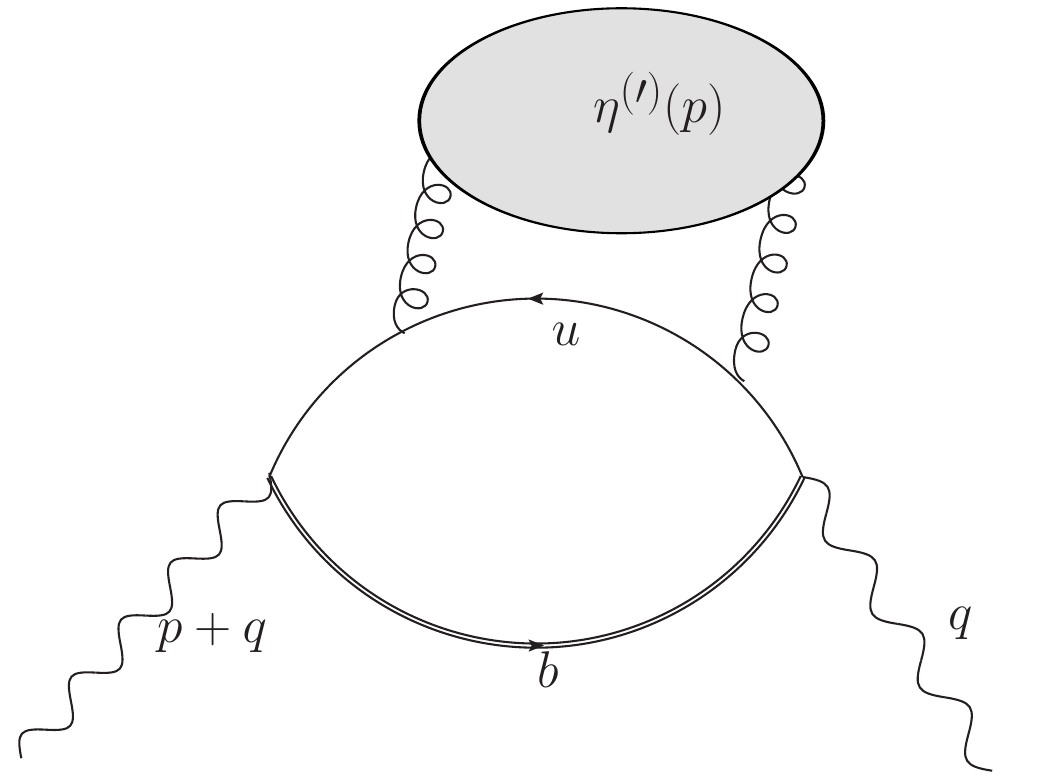}\hspace{1cm}
\includegraphics[width=3.5cm]{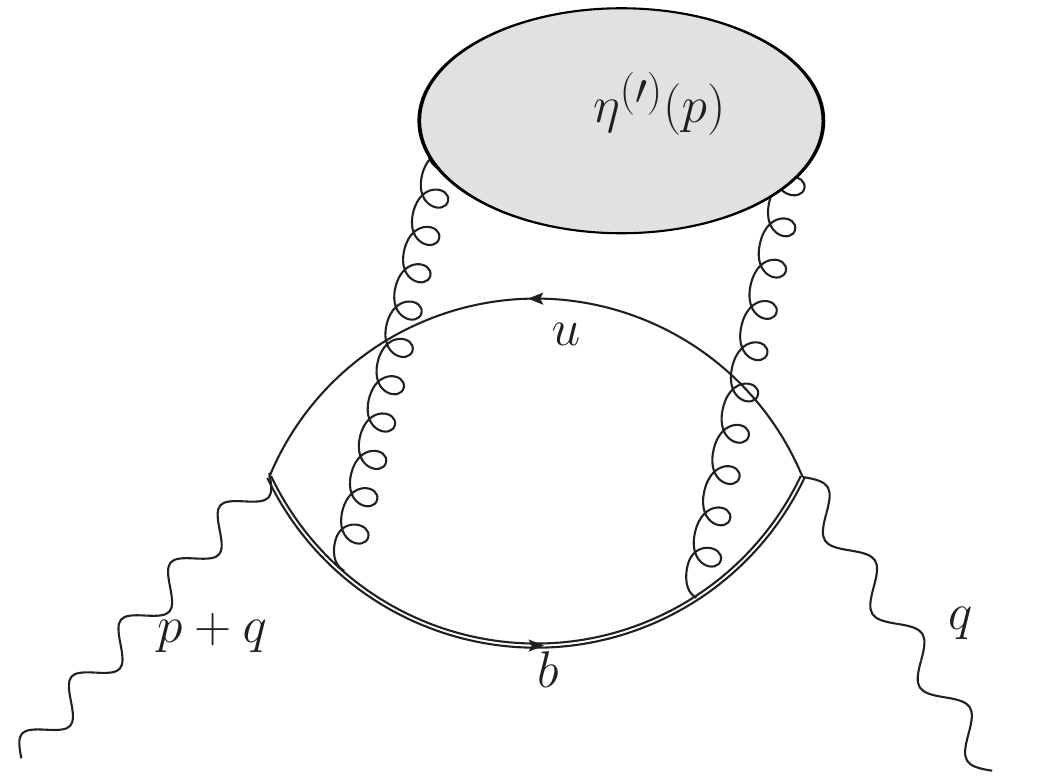}\hspace{1cm}
\includegraphics[width=3.5cm]{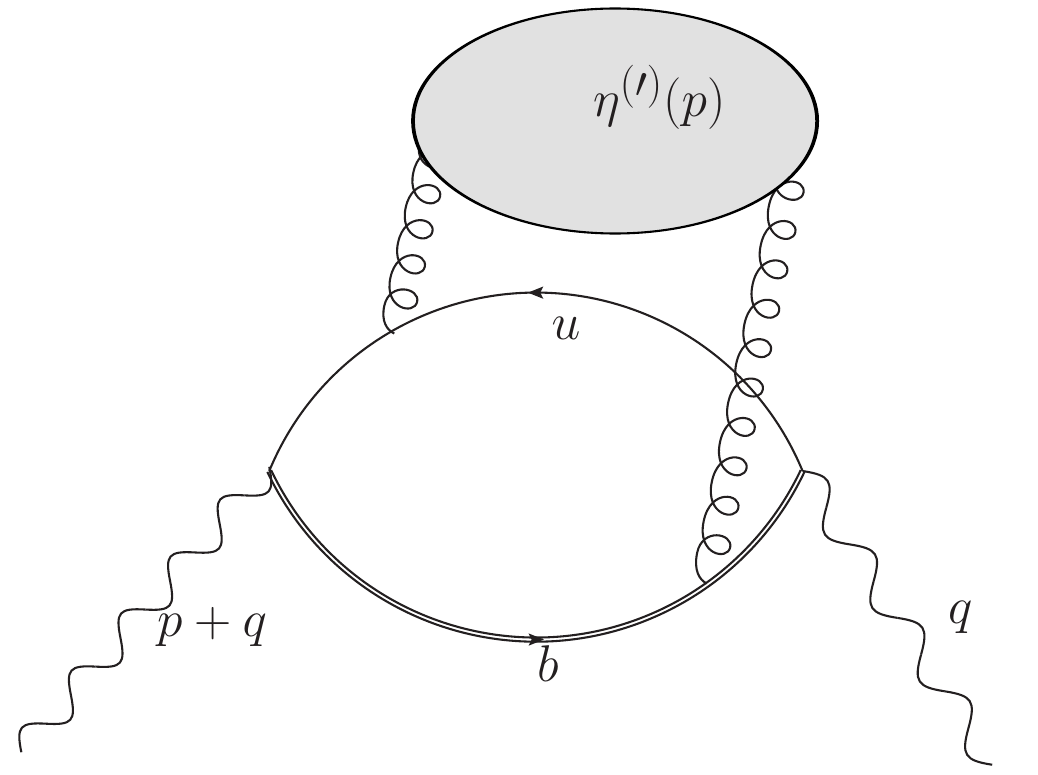}\\
\caption{ \it Diagrams contributing to the gluonic hard-scattering amplitudes at $O(\alpha_s)$.The first diagram is IR divergent and its divergence will be absorbed by the evolution of the gluon DA. See text.}
\label{fig-gluons}
%\end{center}
\end{figure}
%}
Due to the specific properties of $\eta$ and $\eta^\prime$ mesons discussed above, there are additional gluonic diagrams contributing to $M \to \eta^{(\prime)}$ form factors shown in Fig.3. 
These contribution has only been calculated for $f^+_{B\eta^{(\prime)}}$ form factor at twist-2 level in \cite{BallJones} and for $m^2_{\eta^{(\prime)}} = 0$. Here we are going to calculate these contributions for other form factors $f^0_{M\eta^{(\prime)}}$ and $f^T_{M\eta^{(\prime)}}$ by neglecting $O(\alpha_s m_{\eta^{(\prime)}})$ effects in both DAs and the hard-scattering part. This approximation is justified having in mind that parameters of DA for the gluonic DA of $\eta$ and $\eta^\prime$ are badly known, see the values of $b_2^{\eta^{(\prime)} ,g}$ parameter below. 

By using hadronic dispersion relation in the virtuality $(p+q)^2$ of the current in the $B$ channel, we can relate the 
correlation function (\ref{eq:corr})
to the $B \to \eta^{(\prime)}$ matrix elements, 
\be
\langle \eta^{(\prime)}(p)|\bar{u} \gamma_\mu b |\bar B(p+q)\rangle=
2f^+_{B\eta^{(\prime)}}(q^2)p_\mu +\left(f^+_{B\eta^{(\prime)}}(q^2)+f^-_{B\eta^{(\prime)}}(q^2)\right)q_\mu\,,
\label{eq:fplBpi}
\ee
\be
\langle \eta^{(\prime)}(p)|\bar{u} \sigma_{\mu \nu}q^\nu b
|\bar B(p+q)\rangle=
\Big [q^2(2p_\mu+q_\mu) - (m_B^2-m_{\eta^{(\prime)}}^2) q_\mu\Big ]
\frac{i f_{B\eta^{(\prime)}}^T(q^2)}{m_B+m_{\eta^{(\prime)}}}\,.
\label{eq:fsigBpi}
\ee
and extract the form factors. In the literature it sometimes appears that the form factors are defined as above by divided by a factor $\sqrt{2}$ to match the transition form factors of $\eta,\eta^\prime$ with those of a pion when there is no $\eta-\eta^\prime$ mixing and in the limit of the conserved SU(3)-flavour symmetry \cite{BallJones}. 
     
Inserting hadronic states with the $B$-meson quantum numbers between the currents
in (\ref{eq:corr}), and isolating the ground-state $B$-meson contributions 
for all three invariant amplitudes $F(q^2,(p+q)^2)$, $\widetilde{F}(q^2,(p+q)^2)$ and $F^T(q^2,(p+q)^2)$ and 
using (\ref{eq:fplBpi}) and (\ref{eq:fsigBpi}) obtains:
\ba
\hspace*{-2cm} && f^+_{B\eta^{(\prime)}}(q^2) = \frac{e^{m_B^2/M^2}}{2m_B^2 f_B}
\Bigg[F_0(q^2,M^2,s_0^B)+
\frac{\alpha_s C_F}{4\pi}\left ( F_1(q^2,M^2,s_0^B) +  F_1^{gg,+}(q^2,M^2,s_0^B) \right )
\Bigg]\,,
\nonumber \\
\label{eq:fplusLCSR}
\\
&& f^+_{B\eta^{(\prime)}}(q^2)+f^-_{B\eta^{(\prime)}}(q^2) = 
\frac{e^{m_B^2/M^2}}{m_B^2 f_B}
\Bigg[\widetilde{F}_0(q^2,M^2,s_0^B)+
\frac{\alpha_s C_F}{4\pi}\widetilde{F}_1(q^2,M^2,s_0^B)
\Bigg]\,,
\label{eq:fplminLCSR}
\\
\hspace*{-1cm}&& f^T_{B\eta^{(\prime)}}(q^2) =
\frac{(m_B+m_\eta^{(\prime)})e^{m_B^2/M^2}}{2m_B^2 f_B}
\Bigg[F^T_{0}(q^2,M^2,s_0^B)
\nonumber \\
&& \hspace*{3cm} + \frac{\alpha_s C_F}{4\pi} \left ( F^T_{1}(q^2,M^2,s_0^B) +  F_1^{gg,T}(q^2,M^2,s_0^B) \right )
\Bigg]\,.
\label{eq:fTLCSR}
\ea
The scalar $B \to \eta^{(\prime)}$ form factor is then a combination of the vector form factor (\ref{eq:fplusLCSR}) and 
the form factor from (\ref{eq:fplminLCSR}), 
\be
f^0_{B\eta^{(\prime)}}(q^2) = f^+_{B\eta^{(\prime)}}(q^2) + \frac{q^2}{m_B^2-m_{\eta^{(\prime)}}^2} f^-(q^2) 
\label{eq:f0}
\ee
and is only present in the semileptonic $B_{(s)}, D_{(s)} \to \eta^{(\prime)} l \nu $  decays when the lepton mass is not neglected and the rare $B_{s}, D_{s} \to \eta^{(\prime)} l^+ l^- $  decays. 
In above, $F_{0(1)}$ and $\widetilde{F}_{0(1)}$ represent the LO (NLO) contributions and  
$f_B=\langle \bar{B}_d |m_b\bar{b}i\gamma_5 d |0 \rangle/m_B^2$ is the $B$-meson decay constant. $F_1^{gg,(+,T)}$ are leading order twist-2 two-gluon contributions calculated explicitly in the paper. At the leading twist-2 level there is no gluonic contribution in (\ref{eq:fplminLCSR}). However, note from (\ref{eq:f0}) that this does not mean that twist-2 two-gluon contributions will not appear in the scalar $f^0_{M\eta^{(\prime)}}$ form factors (\ref{eq:f0}). 

As usual, the quark-hadron 
duality is used to approximate heavier state contribution by introducing the effective threshold parameter $s_0^B$ and 
the ground state contribution of $B$ meson is enhanced by 
the Borel-transformation in the variable $(p+q)^2 \to M^2$. 
Completely analogous relations are valid for $B_s \to \eta^{(\prime)}$ form factors, with the replacement $u \rightarrow s$ 
in (\ref{eq:fplBpi}) and (\ref{eq:fsigBpi}) and by replacing $m_B$ by $m_{B_s}$, $f_B$ by $f_{B_s}$, as well as $M^2$ by $M_s^2$ and 
$s_0^B$ by $s_0^{B_s}$ in (\ref{eq:fplusLCSR} - \ref{eq:fTLCSR}). 
In addition, in the derivation of above expressions for $B_s$, one has to take into account that 
$\langle B_s | \bar{b}i\gamma_5 s |0 \rangle/m_{B_s}^2 = f_{B_s}/(m_b + m_s)$.  
The same is valid for $D,D_s$ form factors with the replacement $m_b \rightarrow m_c$ and the appropriate exchanges described before.

The calculation will be performed in $\overline{MS}$ scheme.  
The $B, B_s$ and $D,D_s$ decay constants will be calculated in the $\overline{MS}$ scheme using the sum rule expressions 
from  \cite{JL} with $O(\alpha_s, m_s^2)$ accuracy. In that way we achieve the consistency of the calculation and the cancellation of uncertainties in the sum rule parameters.    

Each form factor can be written in the form of the dispersion relation:
\ba
F(q^2,M_{(s)}^2,s_0^{M}) = \frac{1}{\pi}\int\limits_{m_b^2}^{s_0^{M}}
ds e^{-s/M^2_{(s)}}\,\mbox{Im}_s F(q^2,s) \, , 
\label{eq:dispE}
\ea
where now $s = (p + q)^2$. 
%, and which for the NLO effectively means:
%\ba
%&& F_1(q^2,M_{(s)}^2,s_0^{B_{(s)}}) =  \frac{f_K}{\pi} \int\limits_{m_b^2}^{s_0^{B_{(s)}}}ds e^{-s/M^2_{(s)}}
%\int_0^1 du\Bigg\{\mbox{Im}_s T_1(q^2,s,u)\,\varphi_K(u)\
%\nonumber\\
%&& \qquad\qquad\qquad +
%\frac{\mu_K}{m_b}\Big[
%\,\mbox{Im}_sT_1^p(q^2,s,u)\,\phi^p_{3K}(u)\,+
%\,\mbox{Im}_sT_1^\sigma(q^2,s,u)\,\phi^\sigma_{3K}(u)\Big]\Bigg\}\,,
%\label{eq:Imconvol}
%

The leading order parts of the LCSR for $f^+_{M\eta^{(\prime)}}$, $f^+_{M\eta^{(\prime)}}+f^-_{M\eta^{(\prime)}}$ and $f^T_{M\eta^{(\prime)}}$ form factors are given in Appendix A.

Up to now, SU(3)-violating effects for $f_{D_{(s)}\eta^{(\prime)}}$, $f_{B_{(s)}\eta^{(\prime)}}$ form factors were not systematically studied, since the effects of inclusion of $m_{\eta^{(\prime)}}^2$ effects complicate the calculation, especially at NLO in the hard-scattering amplitudes. 
%and the effects are expected to be small. 
However, while the complete SU(3)-symmetry breaking corrections in $\eta^{(\prime)}$ DAs of twist-3 and twist-4 are now known \cite{Braun2014}, it is worth to have a consistent picture of all SU(3)-breaking corrections and we will include complete SU(3)-breaking effects in both DAs, as well as in the hard scattering amplitudes at LO. At NLO in the hard-scattering amplitudes, for the cases when the mass of a light quark cannot be neglected, as for $m_s$,  the inclusion of $m_s$ and $m_{\eta^{(\prime)}}^2$ effects complicate the calculation. As already known from the analysis of $B_{(s)} \to K$ from factors done in \cite{DupliJa}, inclusion of quark mass effects leads to the mixing between different twists and the fully consistent calculation with $m_s$ included in the quark propagators is not possible, see discussion in \cite{DupliJa}. 
However, here we have $\eta^\prime$ as a finite-state particle which mass is much larger than $m_s$ and  
therefore, in the NLO quark and gluonic amplitudes we set $m_s = 0$ and $p^2=m_{\eta^{\prime}}^2 \neq 0$.

Each form factor can be expressed as 
\ba
\left ( \begin{matrix}
 f_{B\eta}^{+,0,T}\\
  f_{B\eta^{\prime}}^{+,0,T} \\
 \end{matrix}
 \right )  = U(\phi) \left ( \begin{matrix}
 f_{B\eta^{q}}^{+,0,T}  \\
f_{B\eta^{s}}^{+,0,T} \\
 \end{matrix}  \right ) \,,
\ea
and 
\ba
f_{B\eta^{q}}^{+,0,T}  = f_{B\eta^{q}}^{(\bar{q}q)\,+,0,T} + f_{B\eta^{q}}^{(gg)\,+,0,T} \,, \quad 
f_{B\eta^{s}}^{+,0,T}  = f_{B\eta^{s}}^{(gg)\,+,0,T}
\ea
and explicitly
\ba
f_{B\eta}^{+,0,T} = \frac{f_\eta^{(q)}}{\sqrt{2}} \left ( F_0^{\bar{q}q} +  F_1^{\bar{q}q}  \right ) + f_\eta^1 F_1^{gg,+,0,T} \,,
\nonumber \\
f_{B\eta^\prime}^{+,0,T} = \frac{f_{\eta^\prime}^{(q)}}{\sqrt{2}} \left ( F_0^{\bar{q}q} +  F_1^{\bar{q}q}  \right ) + f_{\eta^\prime}^1 F_1^{gg,+,0,T} \,,
\nonumber \\
f_{B_s\eta}^{+,0,T} =f_\eta^{(s)} \left ( F_0^{\bar{s}s} +  F_1^{\bar{s}s}  \right ) + f_\eta^1 F_1^{gg,+,0,T} \,,
\nonumber \\
f_{B_s\eta^\prime}^{+,0,T} = f_{\eta^\prime}^{(s)}\left ( F_0^{\bar{s}s} +  F_1^{\bar{s}s}  \right ) + f_{\eta^\prime}^1 F_1^{gg,+,0,T} \,,
\label{eq:fftotal}
\ea
where $F_0^{\bar{q}q}$ and $F_0^{\bar{s}s}$ ($F_1^{\bar{q}q}$ and $F_1^{\bar{s}s}$) are LO (NLO) contributions from quark hard-scattering amplitudes for each of the form factors and $F_1^{gg}$ is the NLO gluonic contribution proportional to the singlet-flavour decay constants
\ba
f_\eta^1 &=& \frac{1}{\sqrt{3}} \left ( \sqrt{2} \cos\phi f_q - \sin\phi f_s \right )  \,,
\nonumber \\
f_{\eta^\prime}^1 &=& \frac{1}{\sqrt{3}} \left ( \sqrt{2} \sin\phi f_q + \cos\phi f_s \right ) \,. 
\ea
The $f_{\eta^(\prime)}^{(r)}$ decay constants are given in (\ref{eq:fetaq}). 
Analogous expressions are valid for $D_{(s)} \to \eta^{(\prime)}$ decays.
 
Obviously, for $B,D \to \eta^{(\prime)}$ transitions the main contribution comes from $\eta_q$ meson states and $\eta_s$ contributes only through suppressed  gluonic contributions, while for $B_s,D_s \to \eta^{(\prime)}$ transitions the leading $\eta_s$ meson state contribution will receive, through the gluonic diagrams, a small mixture with $\eta_q$ state.  Also, implicitly there will be mixing with among twist-2 quark and gluonic distribution amplitudes Eq.(\ref{eq:evol}), which will bring $b_2^{\eta^{(\prime),g}}$ dependence in the twist-2 quark LO ($F_0^{\bar{q}q}$ and $F_0^{\bar{s}s}$ ) and NLO contributions  ($F_1^{\bar{q}q}$ and $F_1^{\bar{s}s}$) and $a_2^{\eta^{(\prime),1}}$ dependence to the gluonic contributions $F_1^{gg}$.

%------------------------------------------------------------------

Since $\eta$ and $\eta^{\prime}$ are mixtures of the $|\bar{q} q \rangle$,  $|\bar{s} s \rangle$ in the calculation of the quark contributions we will use (with appropriate substitutions) our NLO results for the hard-scattering part for $B \to \pi$  \cite{DKMMO} and $B_{(s)} \to K$ form factors \cite{DupliJa} with the $p^2$ effects included at the LO (up to twist-4) and NLO level (up to twist-3) and will imply recently derived DAs of $\eta$ and $\eta^{\prime}$ with the SU(3)-breaking effects and the axial anomaly contributions included up to twist-4. 
The gluonic contributions, which are already NLO effect, will be calculated for $p^2 = m_{\eta^{(\prime)}}^2 = 0$.

\section{LCSR for gluonic contributions to the form factors and consistent treatment of the IR divergences appearing} 

The gluonic contributions at the $O(\alpha_s)$ to the $B(D) \to \eta^{(\prime)}$ and $B_s(D_s) \to \eta^{(\prime)}$ form factors come from the diagrams in Fig.3.  The results for the form factors $f_{M\eta^{(\prime)}}^{+,0,T}$ are presented in subsection 4.1. They are added to the quark contributions (\ref{eq:fftotal}) to get the complete result at the order $O(\alpha_s)$. 

The first diagram is Fig.3 is IR (collinear) divergent. This divergence has to disappear for the 
general collinear factorization formula used here 
\be
F(q^2, (p+q)^2) =  \sum_n T_{H}^{(n)}(u,q^{2},(p+q)^2,\mu_F)  \, \otimes \, \Phi_{n,P}(u,\mu_F)  \,, \quad \otimes = \int_0^1 du  \, .  
\label{eq:fact1}
\ee
be valid. The scale $\mu_F$ is the factorization scale. At the twist $n=2$ level, as already mentioned, there will be mixing of quark and gluonic contributions and the hard-scattering (perturbative part) $T_H^{(2)}$ and the distribution amplitude $\Phi_2$ can be represented as 
\ba
T_H^{(2)} = \left ( \begin{matrix} T_{q\bar q} \\ T_{gg} \end {matrix}\right )\,, \quad  
\Phi_{2,P} = \left ( \begin{matrix} \phi_{2,P}^q \\ \phi_{2,P}^q \end {matrix} \right )\, .
\ea
In order to consistently treat this mixing we have to examine the evolution of the DAs, at the same $O(\alpha_s)$ as the calculation of the perturbative part $T_H$. Due to the mixing the standard 
Brodsky-Lepage (BL) evolution equation \cite{ERBL}
\begin{equation}
  \mu_F \frac{\partial}{\partial \mu_F} \psi(u,\mu_F)   =
   V(u,v,\mu_F) \, \otimes \, \psi(v,\mu_F)
         \, ,
\label{eq:eveq}
\end{equation}
will be a matrix equation now, where $V(u,v,\mu_F)$ is the perturbatively calculable evolution kernel
\begin{eqnarray}
V(u,v,\mu_F) &= & 
       \frac{\alpha_S(\mu_F)}{4 \pi} \, V_1(u,v) +
                 \frac{\alpha_S^2(\mu_F)}{(4 \pi)^2}  V_2(u,v) +
                 \cdots \, , 
\label{eq:kernel}
\end{eqnarray}
with the LO kernel of our interest of the form
\ba
V_1(u,v) &= & \left ( \begin{matrix} V_{qq} && V_{qg} \\ V_{gq} && V_{gg} \end{matrix} \right )
\ea
and $V_{ij}$ are well-know evolution kernels \cite{Terentev,baier,KrollPassek} which we cite here for convenience:
\ba
V_{qq}(u,v) &=& 2 C_F \left \{ \frac{u}{v} \left [ 1 + \frac{1}{v-u} \right ] \Theta  (v-u) + 
\left ( \begin{matrix} u \to 1- u \\ v \to 1-v \end{matrix} \right ) \right \}_+ \,,
\nonumber \\
V_{qg}(u,v) &=& -2 \sqrt{n_f C_F} \left \{ \frac{u}{v^2}  \Theta  (v-u) - 
\left ( \begin{matrix} u \to 1- u \\ v \to 1-v \end{matrix} \right ) \right \} \,, 
\nonumber \\
V_{gq}(u,v) &=& 2 \sqrt{n_f C_F} \left \{ \frac{u^2}{v}  \Theta  (v-u) - 
\left ( \begin{matrix} u \to 1- u \\ v \to 1-v \end{matrix} \right ) \right \} \,,
\nonumber \\
V_{gg}(u,v) &=& 2 N_c \left \{ \frac{u}{v}  \left [ \left ( \frac{\Theta  (v-u)}{v-u} \right )_+  +
\frac{2 u -1}{v} \Theta (v-u) \right ]  + 
\left ( \begin{matrix} u \to 1- u \\ v \to 1-v \end{matrix} \right ) \right \} + \beta_0 \delta (u-v) \,.
\nonumber \\
\ea
These evolution kernels are exactly those which govern the renormalization of the DAs
\ba
\Phi(u) = Z_{\phi, ren} (u,v,\mu_R^2) \, \otimes \, \Phi (v,\mu_R^2 ) \,.
\ea
The connection between $Z$ and the evolution kernel $V$ is given as
\be
  V({\mu}_R^2) = -Z_{\phi, ren}^{-1}({\mu}_R^2) 
          \, \left( {\mu}_R^2
       \frac{\partial}{\partial {\mu}_R^2} Z_{\phi, ren}({\mu}_R^2)
              \right)
          \, , 
\nonumber \\
Z_{\phi, ren}({\mu}_R^2)  = 1 + \frac{\alpha_S(\mu_R)}{4 \pi} \, \frac{1}{\epsilon} V_1(u,v) + \cdots \,.
\ee 
($D= 4 - 2 \epsilon$). 
On the other hand, by calculating the hard-scattering part $T_H$, owing to the fact that final-state quarks are taken to be massless and on-shell (for the case $p^2 = 0$), the amplitude contains collinear singularities. Since $T_H$ is a finite quantity by definition, collinear singularities have to be subtracted.
Therefore, $T$ factorizes as 
\begin{equation}
    T(u,Q^2) = T_H(v, Q^2, \mu_F) \, \otimes \, Z_{T,col}(v, u; \mu_F) \, ,
\label{eq:TTHZ}
\end{equation}
with collinear singularities being subtracted at the scale 
$\mu_F$ and absorbed into the constant $Z_{T,col}$. 
As usual The UV singularities are removed by the renormalization of the fields
and by the coupling-constant renormalization at the (renormalization) scale $\mu_R$. 
Now, in order that the factorization formula is valid, the following has to be satisfied
\be
Z_{T,col}(u,v; \mu_F) \otimes Z_{\phi, ren}(v,\omega; \mu_F)  = \delta(v-\omega)     \, .
\ee
The divergences of $T(u,Q^2)$ and $\Phi(u)$ in (\ref{eq:fact1}) then cancel and 
at the end we are left with
the finite perturbative expressions  for all form transition factors 
\begin{equation}
    F(q^{2}, (p+q)^2)=
      T_{H}(u,Q^{2},\mu_F)  \, \otimes \,  \Phi(u,\mu_F)\,.
\label{eq:tffcf1}
\end{equation}
It is worth pointing out that
the scale $\mu_F$ representing the boundary between the low- and 
high-energy parts in
(\ref{eq:fact1}) plays the role of the separation scale for
collinear singularities
in $T(u,Q^2)$, on the one hand,
and of the renormalization scale for
UV singularities appearing in the perturbatively calculable part
of the distribution amplitude $\Phi(u)$, on the other hand. 
The general discussion and all details of the proof of the cancellation of the factorization scale dependence in the collinear factorization formula (\ref{eq:fact1}) at all orders of calculation can be found in \cite{MelicMALI,NizicMI}. 

In our case of calculating the heavy-to-light transition form factors $f^{+,0,T}$ we face the following situation. The hard-scattering, perturbatively calculable pieces coming from the diagrams from Fig.2 have UV and infra-red singularities at $O(\alpha_s)$. We have already proven in 
\cite{RucklK,DKMMO,DupliJa} for $B\to \pi$ and $B_{(s)} \to K$ form factors that the IR divergences of the quark contributions at twist 2 level cancel exactly with those  coming from the evolution kernel $V_{qq}$. Here, due to the mixing with the twist 2 gluonic contributions, the convolution of $V_{qg}$ of the $T_H$ LO will exactly   cancel the IR divergence in the first gluonic diagram in Fig.3.
At the twist 3 level of $O(\alpha_s)$ the IR divergences of quark diagrams mutually cancel, as shown before in \cite{DKMMO,DupliJa}. 
This gives the final proof of the collinear factorization formula at the given order for the heavy-to-light $M \to \eta^{\prime}$ transition form factors. 

\subsection{Explicit results for the leading two-gluon contributions to the $f^+$ and $f^T$ form factors in $B, B_{s}\to \eta^{(\prime)}$ and $D, D_{s}\to \eta^{(\prime)}$ transitions}

In the calculation of the gluonic contributions to the form factors we have faced the problem of the consistent treatment of the $\gamma_5$ in the dimensional regularization.  
Leading order for the gluonic amplitude is given by one-loop Feynman diagrams in Fig.3. and we have to deal with IR divergence which is a consequence of having massless quarks propagating through the loops.
In the calculation of the gluonic contributions to the form factors it appears a Levi-Civita tensor in the projector of the twist-2 two-gluon DA (\ref{dodform}) and a single $\gamma_5$ matrix in the trace which are both quantities with well-defined properties only in $D=4$ space-time dimensions. Generalization of these quantities in $D$ dimensions is problematic and different approaches to avoid resulting ambiguities can be found in the literature. Moreover, in our case there is no gluonic contributions which appear at LO on $\alpha_s$ that would greatly help in resolving the $\gamma_5$ problem at NLO level. 
The problem was not addressed in the paper where the gluonic amplitude was evaluated for the first time \cite{BallJones} and it is not clear how they resolved the ambiguities.

In the case of the interest it is possible to completely avoid $\gamma_5$ problem and all connected complications since the IR divergence is direct consequence of the massless quark lines and putting a small mass $m$ in massless quark propagators regularizes (removes) the divergence. 
As a consequence, we are not forced to use dimensional regularization and calculation can be performed in four dimensions without any problem. 
Note that putting mass in quark propagators doesn't spoil any of properties and symmetries of the amplitude contrary to the case
when, so called, mass regularization is used on gluon propagators. 
At the very end of calculation it is necessary to expand final result around zero for the small introduced quark mass $m$. The IR divergence will
now reappear as $\ln (m^2)$ term and it is straightforward to connect it with 1/($D$-4) term in the framework of dimensional regularization.

The obtained expressions are as follows: 
\ba
F_1^{gg, i}(q^2,M^2,s_0^M)  = f_{\eta^{(\prime)}}^1 b_2^{\eta^(\prime),g} \frac{1}{C_F} \int_{m^2}^{s_0^M} 
\exp^{-s/M^2} f^{gg,i}(s,q^2) \,, 
\ea
where the gluonic contribution to $f^+$ form factors is 
\ba
f^{gg,+} &=& 20 m^2 \frac{(s-m^2)}{27 \sqrt{3} (s-q^2)^5} 
\bigg( 3 (m^2-q^2) (5 m^4-5 m^2 (q^2+s)+q^4+3 q^2 s+s^2)  \nonumber \\
&&  \left . \hspace*{-1.5cm}  \left(2 \log \left(\frac{s-m^2}{m^2}\right)-\log
   \left(\frac{\mu ^2}{m^2}\right)\right)
   \right . \nonumber \\
&&\hspace*{-1.5cm} -
    \left(37 m^6-m^4 \left(56 q^2+55 s\right)+m^2 \left(18 q^4+76 q^2 s+17
   s^2\right)+3 q^6-27 q^4 s-11 q^2 s^2-2 s^3\right)
   \bigg)
   \nonumber \\
   \label{eq:fplusg}
\ea
and the corresponding contribution to $f^T$ form factors has the following form 
\ba
&& f^{gg,T} = 5 m \frac{s -m^2}{27\sqrt{3} (s-q^2)^5}
\bigg(   12 q^2 \frac{s}{s -m^2} \left(q^4+3 q^2 s+s^2\right) \log \left(\frac{s}{m^2}\right)
  \nonumber \\
   && \left . + 6 (m^2-q^2) \left(5 m^4-5 m^2
   (q^2+s)+q^4+3 q^2 s+s^2\right) \left(2 \log \left(\frac{s-m^2}{m^2}\right)-\log
   \left(\frac{\mu ^2}{m^2}\right)\right)
\right .  \nonumber \\
   && 
- \left(59 m^6-m^4 (72 q^2+85 s)+m^2 s (84 q^2+23 s)+3
   (6 q^6+6 q^4 s+6 q^2 s^2-s^3)\right)
\bigg)
\nonumber \\
 \label{eq:fTg}
\ea
with $m = m_{c,b}$.

With respect to the fact there is no LO $O(\alpha_s^0)$  twist-2 gluon contributions and following the discussions at the beginning of Sec.4, obviously there is no gluonic contributions to $f^+ + f^-$ form factors at this order of calculation. 

The result for the gluonic $O(\alpha_s)$ contribution to $f^+$ form factors was  first given in the appendix of \cite{BallJones}. Our result (\ref{eq:fplusg}) does not completely agree with the one presented there. While we agree in the part being proportional to the logarithmic terms, there is a disagreement between the coefficients in the second line of (\ref{eq:fplusg}) and the expression (A.1) from \cite{BallJones}. Since those terms are exactly those which change with the different treatment of $\gamma_5$, and the authors of \cite{BallJones} have not placed any comment how they  have resolved the $\gamma_5$ ambiguities in the calculation of $f^{gg,+}$, we assume that the difference comes from the improper treatment of the $\gamma_5$ in \cite{BallJones}.  

The result for the gluonic $O(\alpha_s)$ contribution to $f^T$, Eq.(\ref{eq:fTg}) is a new result. 

%%%%%%%%%%%%%%%%%%%%%%%%%%%%%%%%%%%%%%%%%%%%%%%%%%%%%%%%%%%%%%%%%%%%%%%%%%
\section{Predictions for $B, B_{s}\to \eta^{(\prime)}$ and $D, D_{s}\to \eta^{(\prime)}$ form factors ($f^+, f^0$ and $f^T$) the form factors} 
%%%%%%%%%%%%%%%%%%%%%%%%%%%%%%%%%%%%%%%%%%%%%%%%%%%%%%%%%%%%%%%%%%%%%%%%%%

The prediction for $B, B_{s}\to \eta^{(\prime)}$ and $D, D_{s}\to \eta^{(\prime)}$ form factors ($f^+, f^0$ and $f^T$) the form factors will be given in the $\overline{MS}$ scheme by using the input parameters listed in Appendix B. 

From expressions (\ref{eq:fplminLCSR},\ref{eq:fTLCSR},\ref{eq:f0}) we see that we need the heavy-meson decay constants of $B_{(s)}$ and $D_{(s)}$ in the calculation. As usually done, to achieve partial cancellation of the uncertainties in the calculation the two-point QCD sum rules for the decay constants $f_B, f_{B_s}$ and $f_D, f_{D_s}$ is used in the same scheme, with $O(\alpha_s, m_s^2)$ corrections included \cite{JL}. We have used the same level of accuracy as in the calculation of the form factors, i.e ${\cal O}(\alpha_s)$ in both, the perturbative and nonperturbative (quark condensate) part and in the determination of the sum rules parameters $\overline{s}_0^{B_{(s)}}$ and $\overline{M}_{(s)}^2$ have used the usual consistency conditions in the sum rule calculations. 

The resulting predictions for $f_M$, together with the fitted sum rule parameters for each of the mesons are given in the Appendix B, Tables 2-4. Here we quote the calculated values from Table 4: 
\ba
& & f_D = 191 \pm 9 \; {\rm MeV} , \,  f_{D_s} =  219 \pm 7 \; {\rm MeV} \,,
\nonumber \\
& & f_B = 215 \pm 7 \; {\rm MeV} , \,  f_{B_s} =  246 \pm 8 \; {\rm MeV} \,,
\ea
where the quoted error intervals are coming from the variation of $\overline{s}_0^{M}$ and $\overline{M}_{M}^2$ only since other uncertainties are canceled in ratios in Eqs.(\ref{eq:fplminLCSR},\ref{eq:fTLCSR},\ref{eq:f0}). 
By comparing our results with the previous LCSR results and the most recent determinations from \cite{KhodjamPivovarov}, where in the perturbative part the higher order corrections were included, we see good agreement. The results are also within uncertainties of the lattice QCD calculations of the same decay constants \cite{latticeDC}. 

For the $f_D$ and $f_{D_s}$ the experiment gives somewhat larger values  
\cite{PDG2014}, 
\ba
f_D = 204.6 \pm 5.0 \; {\rm MeV} , \,  f_{D_s} =  257.5 \pm 4.6 \; {\rm MeV} \,, \nonumber
\ea
but still consistent within the complete LCSR error results \cite{KhodjamPivovarov}. 

%\subsection{Updated predictions for the $B_{(s)}\to K$   form factors}

The renormalization scale is given by the expression $\mu_{B_(s)} = \sqrt{m_{B_{(s)}}^2 - m_b^2}$ and similarly for $D_{(s)} \to \eta^{(\prime)}$ transitions. Therefore, for the renormalization scale
 we use $\mu= 3$ GeV, for the $f_{B\eta^{(\prime)}}^{0,+,T}$ form factors and $\mu_s= 3.4$ GeV for $f_{B_s\eta^{(\prime)}}^{0,+,T}$ and for $\mu_{D} = 1.4$ GeV and $\mu_{D_s} = 1.5$ GeV. 
As usual, we will check the sensitivity of the results on the variation of above scales and will include it in the error estimation. 

The method of extraction of the Borel parameters $M$ and the effective thresholds $s_0$ for $f_{M\eta^{(\prime)}}^{+,0,T}$ form factors is the same as described 
in \cite{DKMMO}. It relies on the requirement that the derivative over $-1/M^2$ of the expression of the complete LCSR
 for a particular form factor, which gives heavy-meson masses $m_M^2$,  does not deviate more than $0.5-2.5 \%$ from the experimental values
 for those masses. Additional requirements such as that the subleading twist-4 terms in the LO, are small, less than $10\%$ of the LO twist-2 term, that 
the NLO corrections of twist-2 and twist-3 parts are not exceeding $30\%$ of their LO counterparts, and that the subtracted continuum remains small, are also satisfied. 
These demands provide us the central values for the LCSR parameters listed in Tab.\ref{tab-5}.

The estimated form factors for $B_{(s)} \to \eta^{(\prime)}$ are as follows:
\ba
f^+_{B\eta}(0) &=& 0.168^{+0.041}_{-0.047} = 0.168 \pm 0.003\, (b_2^{\eta,g})\pm  0.002 (s_0,M) \pm ^{0.041}_{0.047}  ({\rm mix}) \pm  ^{0.005}_{0.003} ({\rm rest}) \,,\nonumber \\
f^+_{B\eta^{\prime}}(0) &=& 0.130^{+0.036}_{-0.032} = 0.130 \pm 0.020\, (b_2^{\eta^{\prime},g}) \pm   0.002 (s_0,M) \pm  ^{0.030}_{0.032} ({\rm mix}) \pm  ^{0.005}_{0.002} ({\rm rest}) \,, \nonumber\\
|f^+_{B_s\eta}(0)| &=& 0.212^{+0.015}_{-0.013}  = 0.212 \pm 0.003\, (b_2^{\eta,g}) \pm    0.003 (s_0,M) \pm  0.012 ({\rm mix}) \pm  ^{0.008}_{0.003} ({\rm rest}) \,,\nonumber\\
f^+_{B_s\eta^{\prime}}(0) &=& 0.252^{+0.023}_{-0.020}  = 0.252 \pm 0.019\, (b_2^{\eta^{\prime},g}) \pm   0.004 (s_0,M) \pm  0.005 ({\rm mix}) \pm  ^{0.011}_{0.002} ({\rm rest})  \,,\nonumber\\
\label{eq:results52}
\ea
\ba
f^T_{B\eta}(0) &=& 0.173^{+0.041}_{-0.035} = 0.173 \pm 0.002\, (b_2^{\eta,g}) \pm 0.003 (s_0,M) \pm  ^{0.040}_{0.035} ({\rm mix}) \pm  ^{0.007}_{0.003} ({\rm rest})  \,,\nonumber\\
f^T_{B\eta^{\prime}}(0) &=& 0.141^{+0.032}_{-0.030}  = 0.141 \pm 0.015\, (b_2^{\eta^{\prime},g}) \pm  0.002 (s_0,M) \pm  ^{0.028}_{0.026} ({\rm mix}) \pm  ^{0.006}_{0.003} ({\rm rest}) \,,\nonumber\\
|f^T_{B_s\eta}(0)|&=& 0.225^{+0.019}_{-0.014}  = 0.225 \pm 0.002\, (b_2^{\eta,g}) \pm   0.004 (s_0,M) \pm  ^{0.014}_{0.013} ({\rm mix}) \pm  ^{0.012}_{0.002} ({\rm rest}) \,,\nonumber\\
f^T_{B_s\eta^{\prime}}(0) &=& 0.280^{+0.022}_{-0.016} = 0.280 \pm 0.014\, (b_2^{\eta^{\prime},g}) \pm  0.004 (s_0,M) \pm  ^{0.006}_{0.007} ({\rm mix}) \pm  ^{0.015}_{0.002} ({\rm rest}) \,,\nonumber\\
\ea
and for $D_{(s)} \to \eta^{(\prime)}$:
\ba
f^+_{D\eta}(0) &=& 0.429^{+0.165}_{-0.141}  = 0.429 \pm 0.009\, (b_2^{\eta,g}) \pm    ^{0.004}_{0.001} (s_0,M) \pm  ^{0.164}_{0.141} ({\rm mix}) \pm  ^{0.013}_{0.008} ({\rm rest}) \,,\nonumber\\
f^+_{D\eta^{\prime}}(0) &=& 0.292^{+0.113}_{-0.104} =  0.292 \pm 0.045\, (b_2^{\eta^{\prime},g}) \pm     ^{0.009}_{0.007} (s_0,M) \pm  ^{0.099}_{0.091} ({\rm mix}) \pm  ^{0.015}_{0.011} ({\rm rest})\,,\nonumber\\
|f^+_{D_s\eta}(0)| &=& 0.495^{+0.030}_{-0.029} = 0.495 \pm 0.007\, (b_2^{\eta,g}) \pm  ^{0.004}_{0.002} (s_0,M) \pm  ^{0.027}_{0.024} ({\rm mix}) \pm  ^{0.016}_{0.009} ({\rm rest}) \,,\nonumber\\
f^+_{D_s\eta^{\prime}}(0) &=& 0.557^{+0.048}_{-0.045} = 0.557 \pm 0.041\, (b_2^{\eta^{\prime},g}) \pm ^{0.011}_{0.008} (s_0,M) \pm  ^{0.010}_{0.008} ({\rm mix}) \pm  ^{0.022}_{0.014} ({\rm rest})\,,\nonumber\\
\label{eq:results54}
\ea
\ba
f^T_{D\eta}(0) &=& 0.435^{+0.115}_{-0.107}  = 0.435 \pm 0.008\, (b_2^{\eta,g}) \pm    ^{0.005}_{0.003} (s_0,M) \pm ^{0.112}_{0.106} ({\rm mix}) \pm  ^{0.177}_{0.151} ({\rm rest})\,, \nonumber\\
f^T_{D\eta^{\prime}}(0) &=& 0.337^{+0.118}_{-0.147} =  0.337 \pm 0.055\, (b_2^{\eta^{\prime},g}) \pm ^{0.013}_{0.051} (s_0,M) \pm  ^{0.100}_{0.101} ({\rm mix}) \pm  ^{0.080}_{0.077} ({\rm rest})\,, \nonumber\\
|f^T_{D_s\eta}(0)|&=& 0.441^{+0.091}_{-0.087} = 0.441 \pm 0.007\, (b_2^{\eta,g}) \pm ^{0.052}_{0.005} (s_0,M) \pm  ^{0.030}_{0.031} ({\rm mix}) \pm  ^{0.068}_{0.082} ({\rm rest})\,, \nonumber\\
f^T_{D_s\eta^{\prime}}(0) &=& 0.655^{+0.072}_{-0.065} = 0.655 \pm 0.050\, (b_2^{\eta^{\prime},g}) \pm ^{0.015}_{0.014} (s_0,M) \pm  ^{0.036}_{0.030} ({\rm mix}) \pm  ^{0.034}_{0.026} ({\rm rest})\,.\nonumber\\
\label{eq:results}
\ea

These results are predictions given with $b_2^{\eta^{(\prime),g}} = 0$ and then varied 
within the interval $\Delta b_2^{\eta^{(\prime),g}} =  \pm 20$, which dependence is explicitly displayed in the errors. 
The errors are compilation of the variation of parameters added in quadratures. In the errors we explicitly stress SR parameter dependence $(s_0,\, M)$, $\eta-\eta^{\prime}$ mixing
parameter dependence (mix) and dependences coming from the variation of the rest of parameters (rest=$\{ \mu,\, m_{c,b},\, a_2,\, a_4  \}$).

The errors of the results are much larger for the transitions $B,D \to \eta^{(\prime)}$ where $B,D \to \eta_q$ dominates then for $B_s,D_s \to \eta_s$ decays since the error in the parameter $h_q$ (\ref{eq:hqhs}) is huge, of $O(200\%)$ depending not on $\eta-\eta^{\prime}$ mixing parameters but exhibiting a numerical cancellation among terms. If one would use approximation (\ref{eq:hq-approx}) applied in \cite{Offen} instead, the (rest)-errors would be almost an order of magnitude lower and the mean values would be somewhat larger for those decays, which we assume is the main reason, apart from the rest of $SU(3)_F$ approximations used there, of the discrepancies with some of the results presented in \cite{Offen}. We see that the dominant errors in $M \to \eta^{\prime}$ form factors is coming from the variation of $b_2^{\eta^{(\prime),g}}$ and it amounts to  about $15\%$, while in $M \to \eta$ decays come to $2\%$. Our findings for calculated $B \to \eta^{(\prime)}$ form factors agree very well with those from \cite{BallJones,Li,Aliev1,Aliev2}. 

Their $q$-dependence of the form factors and their ratios is shown in Fig.4-9. 

%%%%%%%%%%%%%%%%%%%%%%%%%%%%%%%%%%%
%%%%%%%%%fig4: all form factors
%\FIGURE[t]{
%%\begin{figure}[t]
%%\begin{center}
%\includegraphics[width=11.5cm]{fig4.eps}
%\caption{ \it The LCSR prediction for
%form factors $f^+_{BK}(q^2)$ (solid line), $f^0_{BK}(q^2)$ (dashed line)
%and $f^T_{BK}(q^2)$ (dash-dotted line)
%at $0<q^2<12$ GeV$^2$ and for $\mu = 3$ GeV, $s_0^{B} = 38\,{\rm GeV}^2$, $M^2 = 18 \, {\rm GeV}^2$ and the central values of all 
%other input parameters. }
%\label{fig-fK}
%%\end{center}
%%\end{figure}
%}
%%%%%%%%%%%%%%%%%%%%%%%%%%%%%%%%%%%%
%%%%%%%%%%%%%%%%%%%%%%%%%%%%%%%%%%%%
%%%%%%%%%%%fig5: all form factors
%\FIGURE[h]{
%%\begin{figure}[t]
%%\begin{center}
%\includegraphics[width=11.5cm]{fig5.eps}
%\caption{ \it The LCSR prediction for
%form factors $f^+_{B_sK}(q^2)$ (solid line), $f^0_{B_sK}(q^2)$ (dashed line)
%and $f^T_{B_sK}(q^2)$ (dash-dotted line)
%at $0<q^2<12$ GeV$^2$ and for $\mu = 3.4$ GeV, $s_0^{B_s} = 39\,{\rm GeV}^2$, $M_s^2 = 19 \, {\rm GeV}^2$ and the central values of all other 
%input parameters. }
%\label{fig-fKs}
%%\end{center}
%%\end{figure}
%}
%%%%%%%%%%%%%%%%%%%%%%%%%%%%%%%%%%%%

%\FIGURE[t]{
\begin{figure}[t]
%\begin{center}
\includegraphics[width=6.5cm]{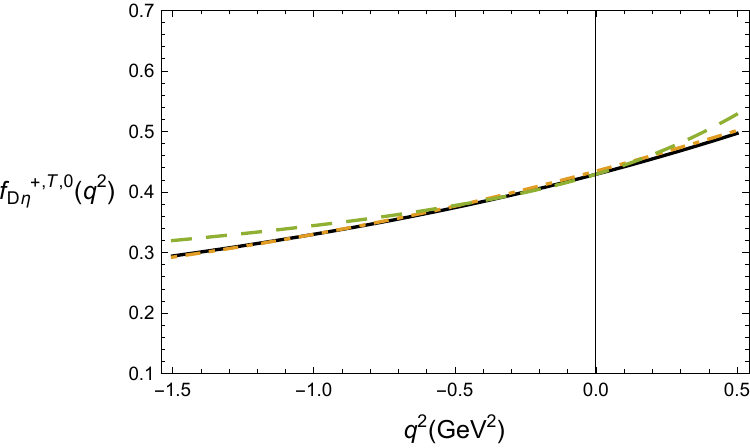}\hspace{1cm}
\includegraphics[width=6.5cm]{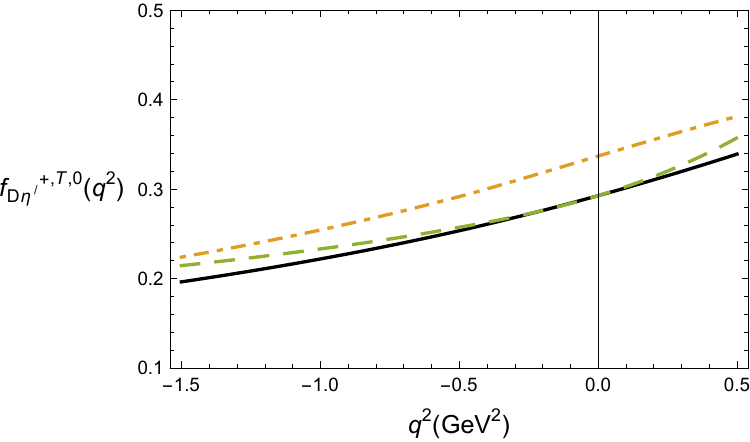}\hspace{1cm}
\\
\includegraphics[width=6.3cm]{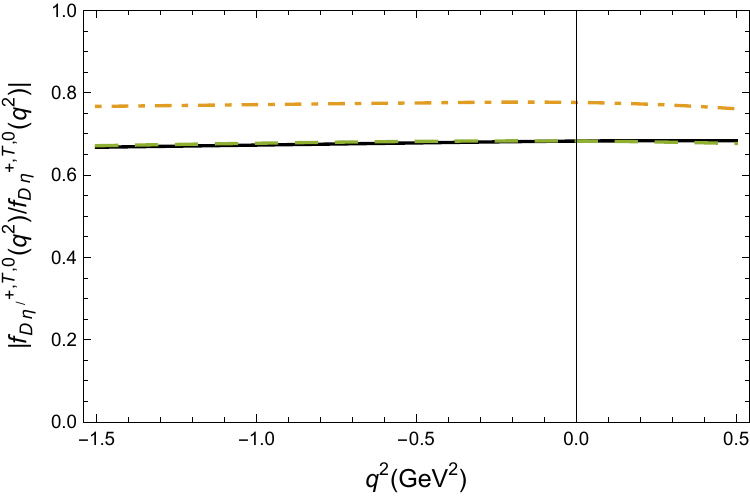}\hspace{1cm}
\\
\includegraphics[width=6.5cm]{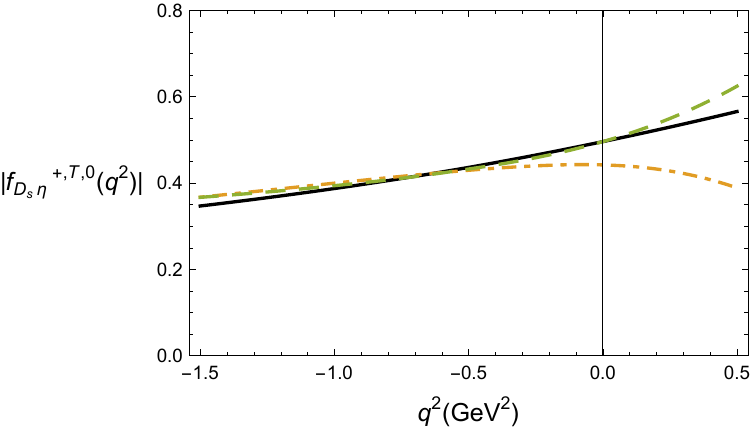}\hspace{1cm}
\includegraphics[width=6.5cm]{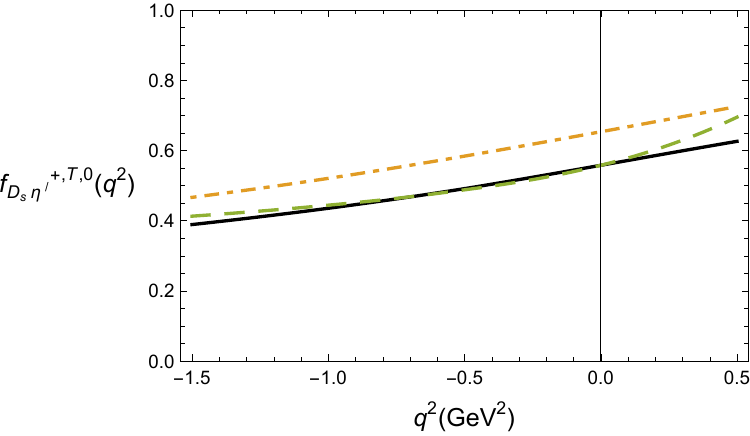}\hspace{1cm}
\\
\includegraphics[width=6.3cm]{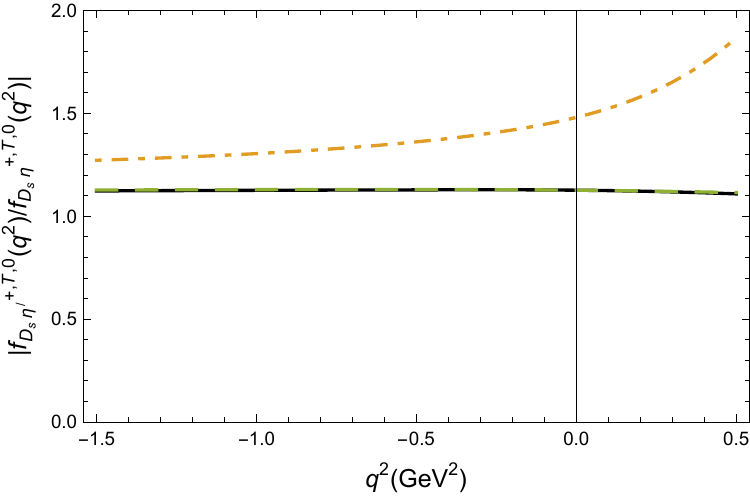}\\
\caption{ \it Form factors for $D_{(s)} \to \eta^{(\prime)}$ decays and their ratios. Solid lines represent $f_{D_{(s)}\eta^{(\prime)}}^+$ form factors, 
dashed-dotted line $f_{D_{(s)}\eta^{(\prime)}}^T$ and dashed line $f_{D_{(s)}\eta^{(\prime)}}^0$ form factors. } 
\label{fig-D}
%\end{center}
\end{figure}
%}
%\FIGURE[h]{
\begin{figure}[t]
%\begin{center}
\includegraphics[width=6.5cm]{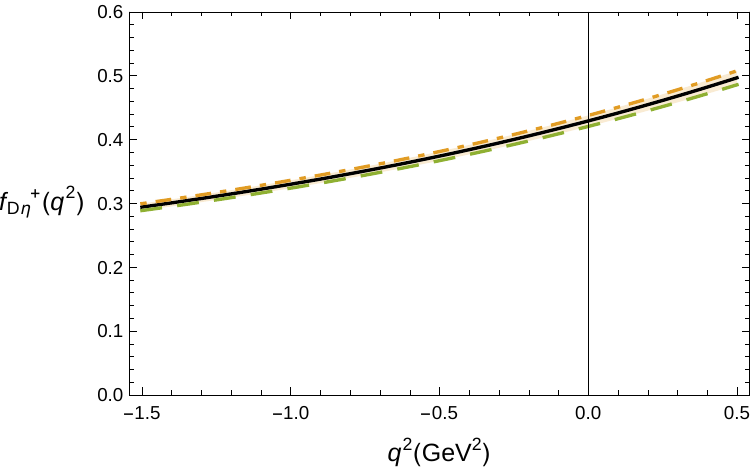}\hspace{1cm}
\includegraphics[width=6.5cm]{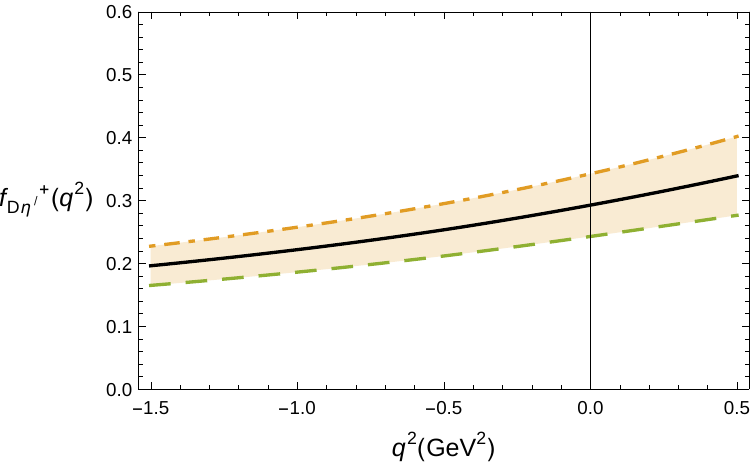}\hspace{1cm}
\\
\includegraphics[width=6.5cm]{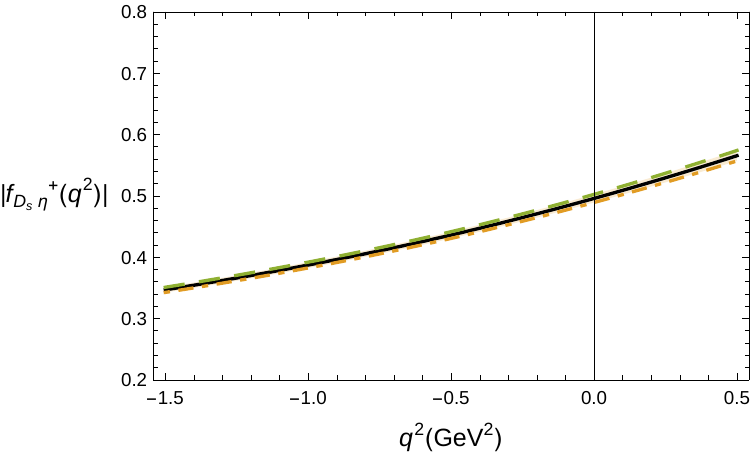}\hspace{1cm}
\includegraphics[width=6.5cm]{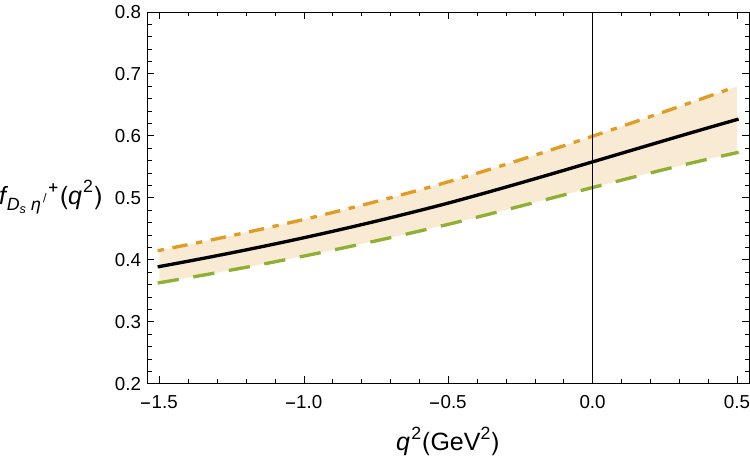}\hspace{1cm}
\caption{ \it Gluonic dependence of $f_{D_{(s)} \eta^{(\prime)}}^+$ form factors. Shaded areas show change of the form factors under the variation of $b_2^{\eta^{(\prime)},g} = 0 \pm 20$. Solid line denotes the result for $b_2^{\eta^{(\prime)},g} = 0$, dashed-dotted for $b_2^{\eta^{(\prime)},g} = 20$ and dashed line for $b_2^{\eta^{(\prime)},g} = -20$.}
\label{fig-Dgluon}
%\end{center}
\end{figure}
%}
%\FIGURE[h]{
\begin{figure}[t]
%\begin{center}
\includegraphics[width=6.5cm]{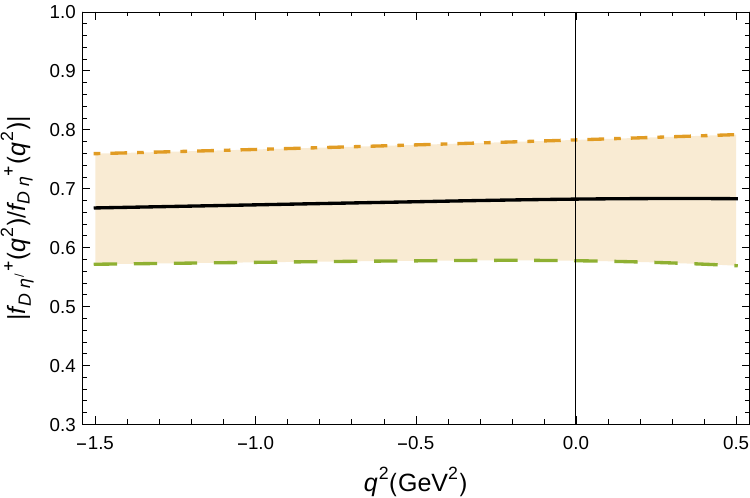}\hspace{1cm}
\includegraphics[width=6.5cm]{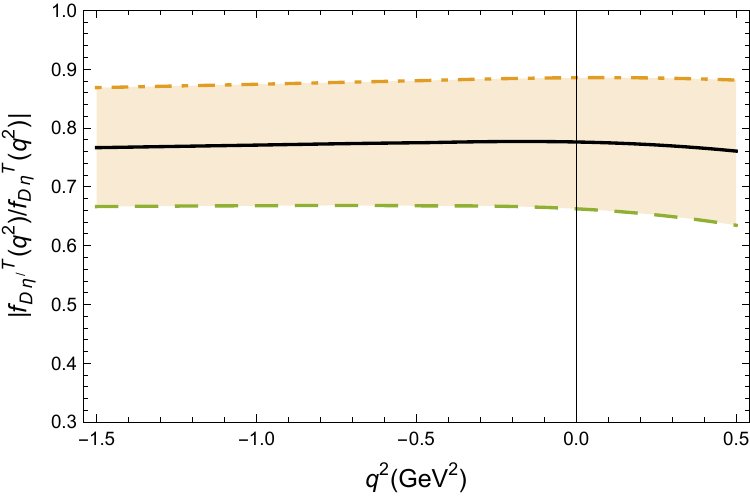}\hspace{1cm}
\\
\includegraphics[width=6.5cm]{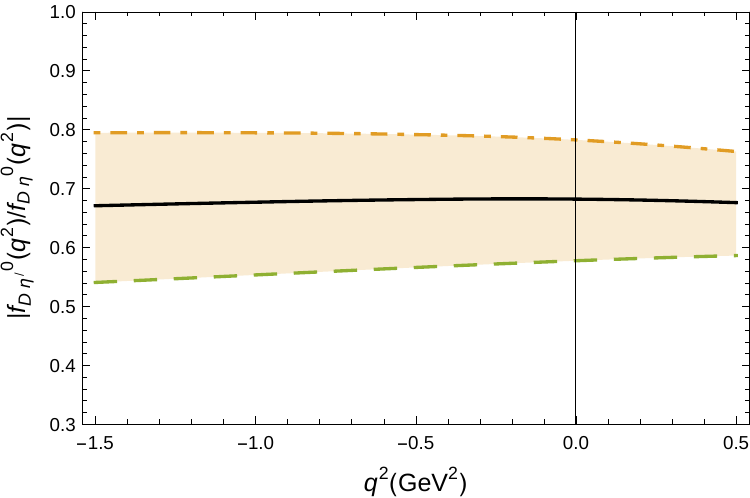}\hspace{1cm}
\\
\includegraphics[width=6.5cm]{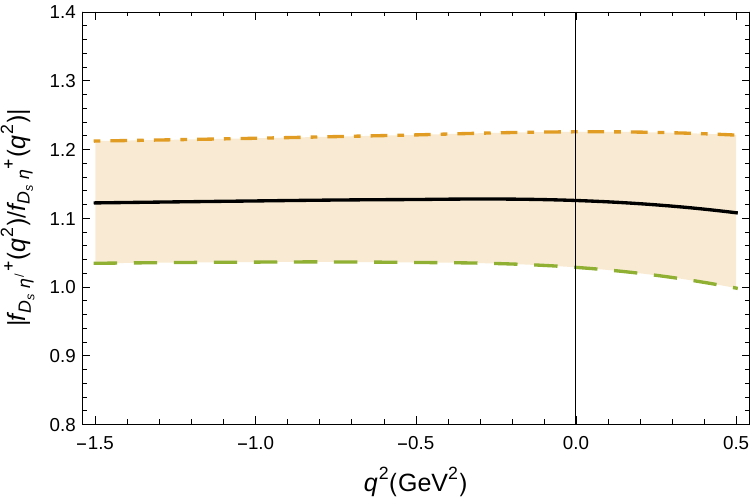}\hspace{1cm}
\includegraphics[width=6.5cm]{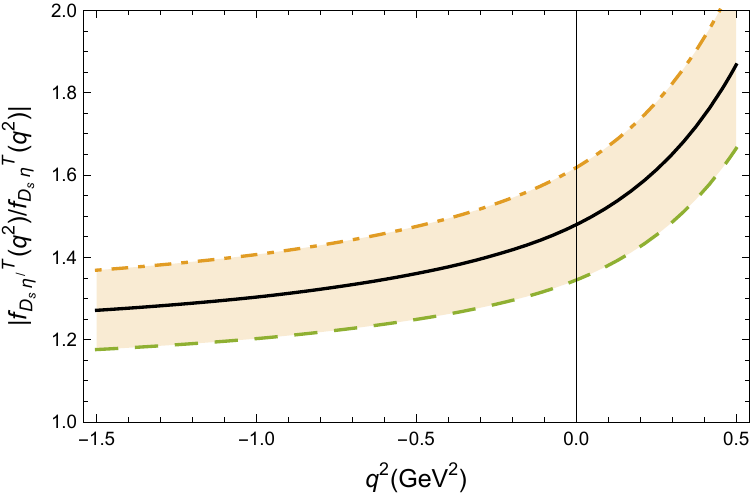}\hspace{1cm}
\\
\includegraphics[width=6.5cm]{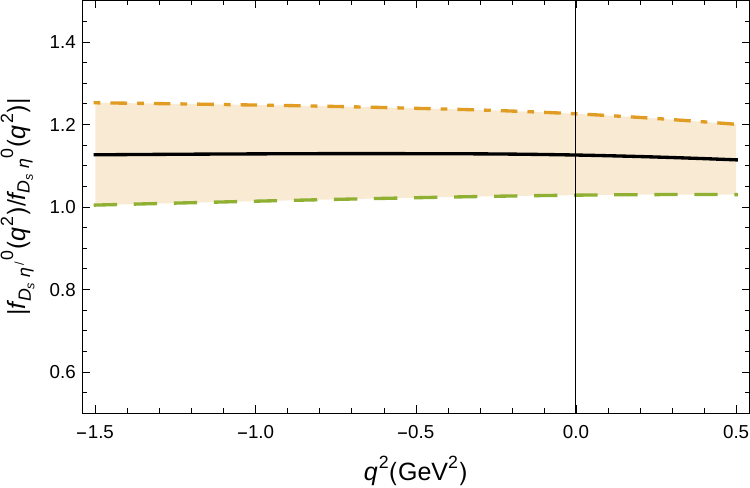}\\
\caption{ \it Gluonic dependence of ratios of $
D_{(s)} \to \eta^{(\prime)}$ form factor ratios. Shaded areas show change of the form factors under the variation of $b_2^{\eta^{(\prime)},g} = 0 \pm 20$. Solid line denotes the result for $b_2^{\eta^{(\prime)},g} = 0$, dashed-dotted for $b_2^{\eta^{(\prime)},g} = 20$ and dashed line for $b_2^{\eta^{(\prime)},g} = -20$.}
\label{fig-DgluonR}
%\end{center}
\end{figure}
%}
%\FIGURE[h]{
\begin{figure}[t]
%\begin{center}
\includegraphics[width=6.5cm]{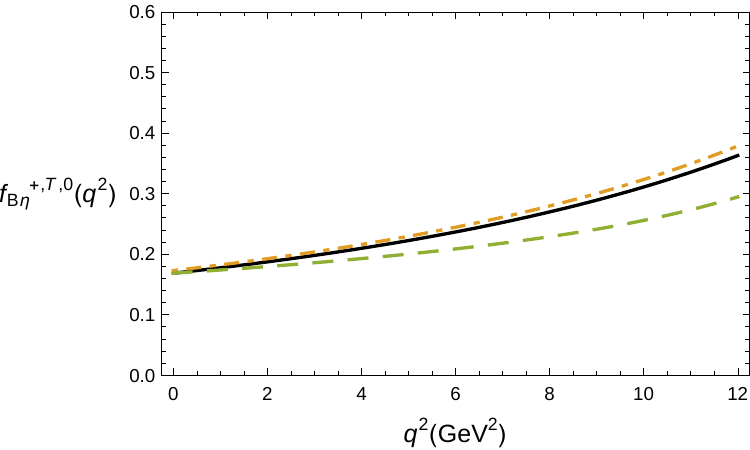}\hspace{1cm}
\includegraphics[width=6.5cm]{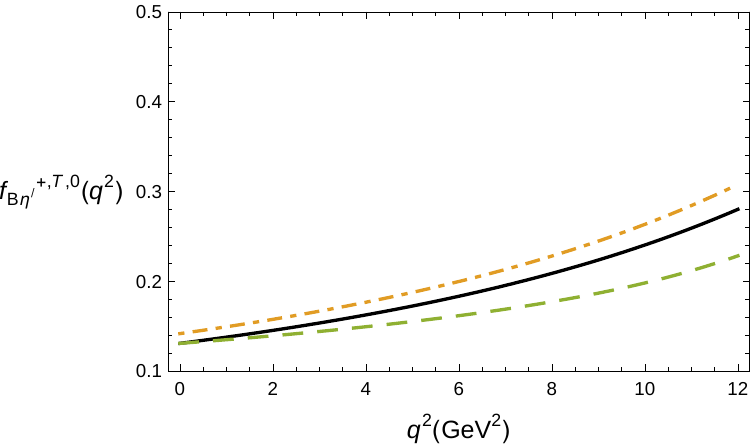}\hspace{1cm}
\\
\includegraphics[width=6.5cm]{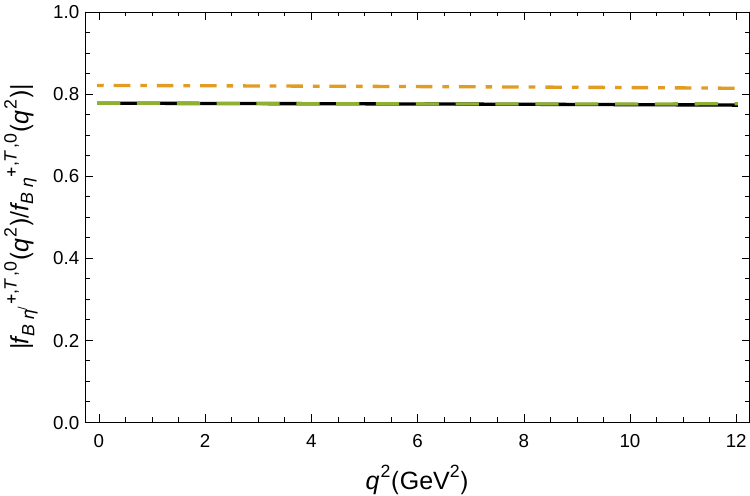}\hspace{1cm}
\\
\includegraphics[width=6.5cm]{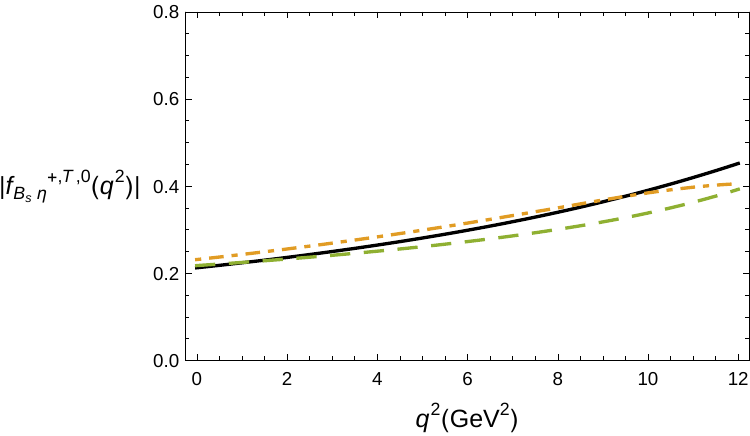}\hspace{1cm}
\includegraphics[width=6.5cm]{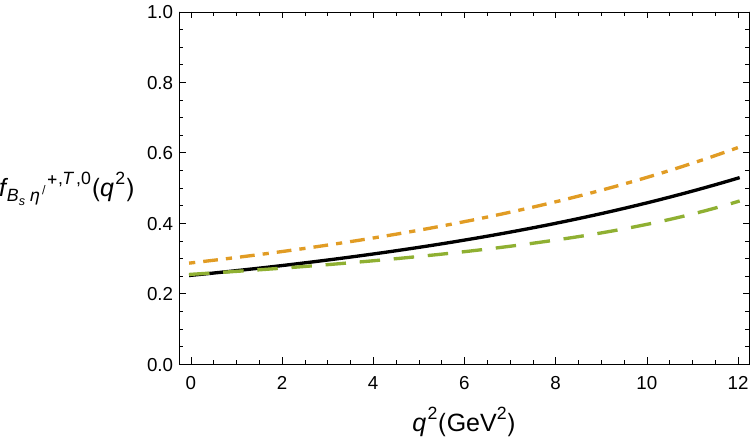}\hspace{1cm}
\\
\includegraphics[width=6.5cm]{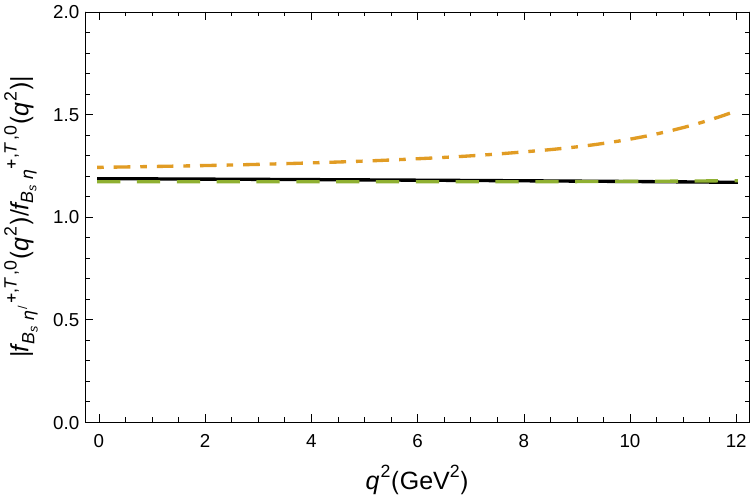}\\
\caption{ \it Form factors for $B_{(s)} \to \eta^{(\prime)}$ decays  and their ratios. Solid lines represent $f_{B_{(s)}\eta^{(\prime)}}^+$ form factors, 
dashed-dotted line $f_{B_{(s)}\eta^{(\prime)}}^T$ and dashed line $f_{B{(s)}\eta^{(\prime)}}^0$ form factors.}
\label{fig-B}
%\end{center}
\end{figure}
%}
%\FIGURE[h]{
\begin{figure}[t]
%\begin{center}
\includegraphics[width=6.5cm]{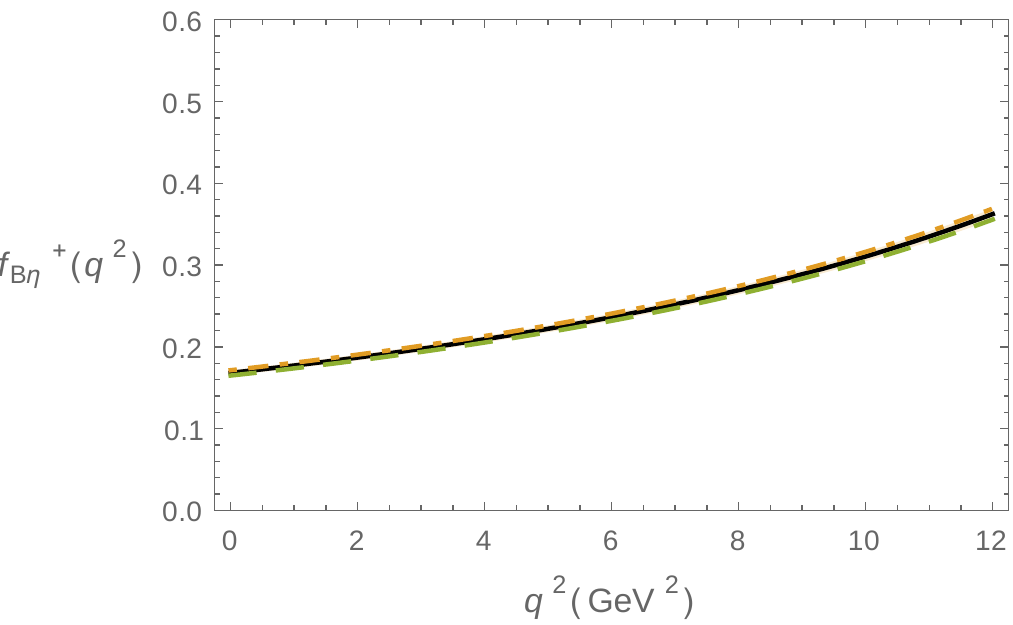}\hspace{1cm}
\includegraphics[width=6.5cm]{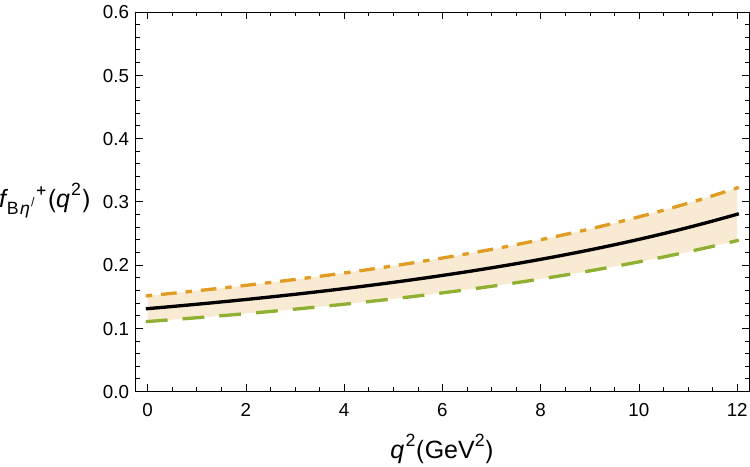}\hspace{1cm}
\\
\includegraphics[width=6.5cm]{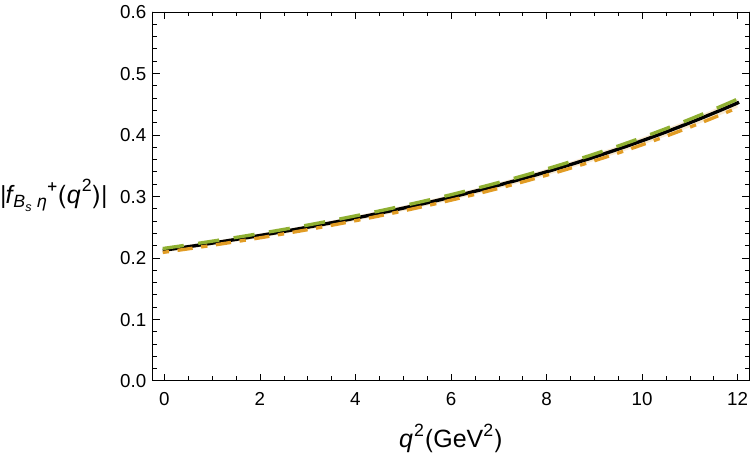}\hspace{1cm}
\includegraphics[width=6.5cm]{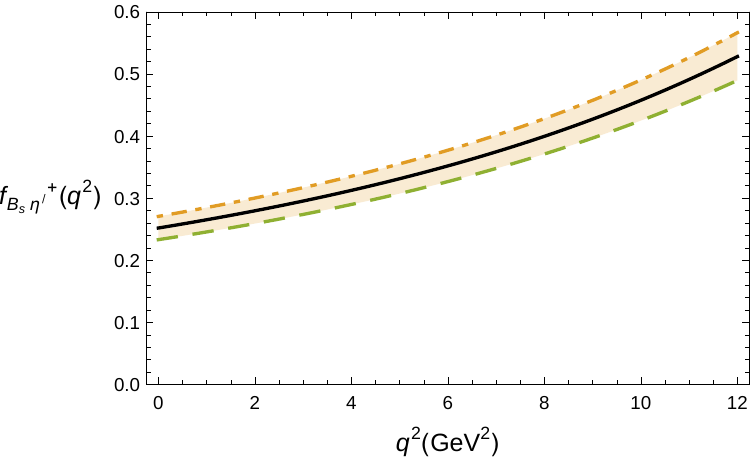}\hspace{1cm}
\caption{ \it Gluonic dependence of $f_{B_{(s)} \eta^{(\prime)}}^+$ form factors. Shaded areas show change of the form factors under the variation of $b_2^{\eta^{(\prime)},g} = 0 \pm 20$. Solid line denotes the result for $b_2^{\eta^{(\prime)},g} = 0$, dashed-dotted for $b_2^{\eta^{(\prime)},g} = 20$ and dashed line for $b_2^{\eta^{(\prime)},g} = -20$.}
\label{fig-Bgluon}
%\end{center}
\end{figure}
%}
%\FIGURE[h]{
\begin{figure}[t]
%\begin{center}
\includegraphics[width=6.5cm]{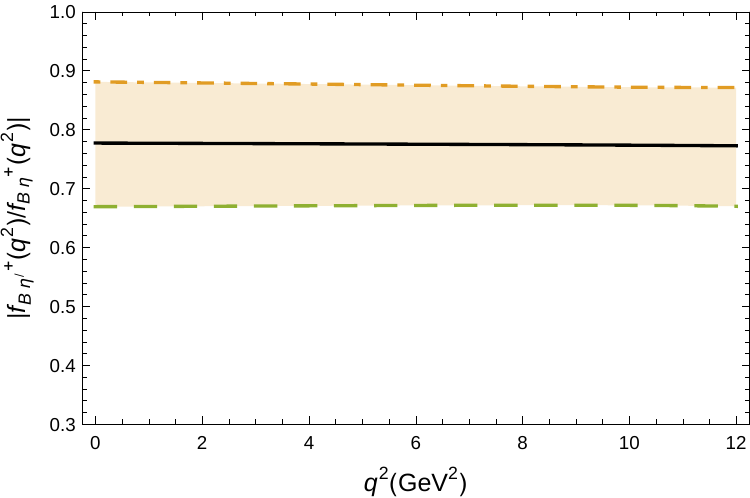}\hspace{1cm}
\includegraphics[width=6.5cm]{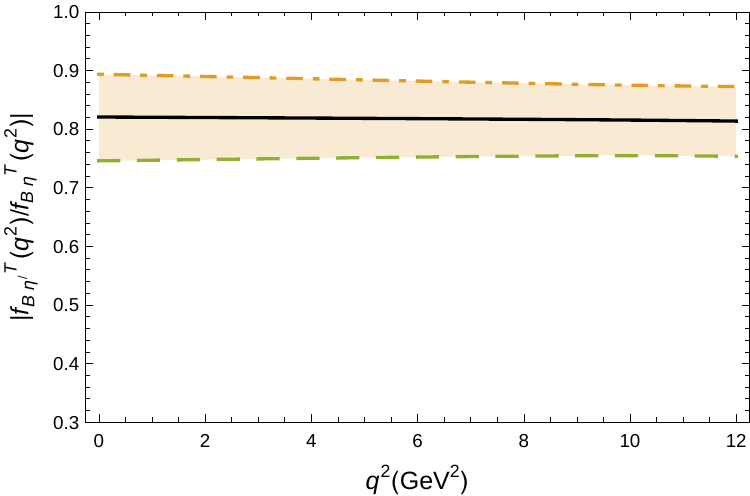}\hspace{1cm}
\\
\includegraphics[width=6.5cm]{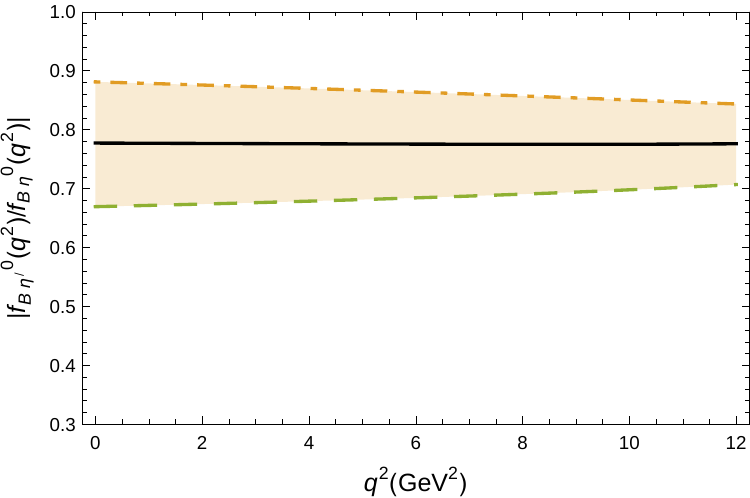}\hspace{1cm}
\\
\includegraphics[width=6.5cm]{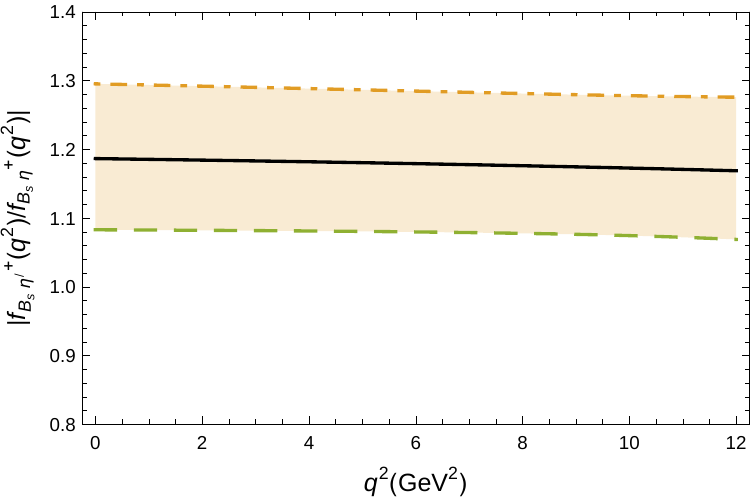}\hspace{1cm}
\includegraphics[width=6.5cm]{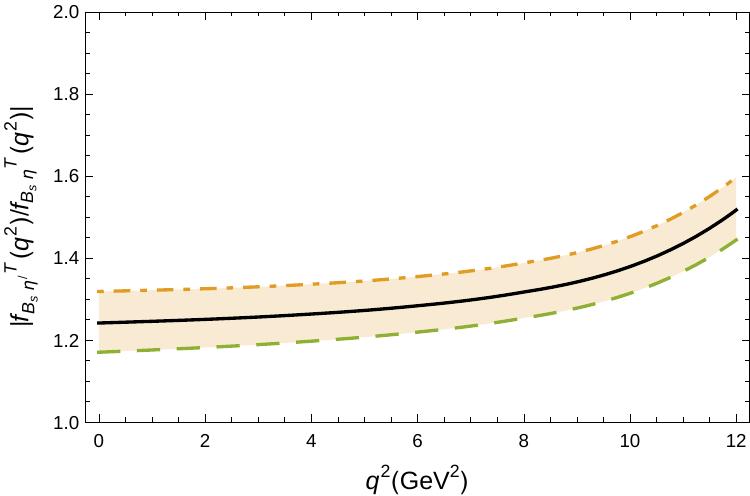}\hspace{1cm}
\\
\includegraphics[width=6.5cm]{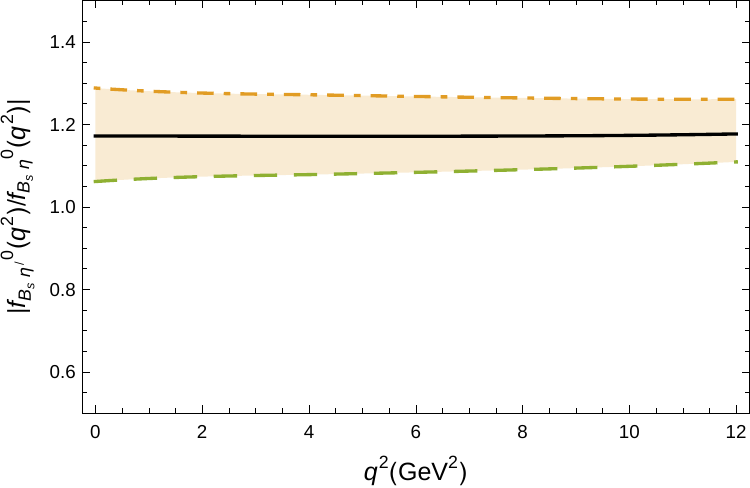}\\
\caption{ \it Gluonic dependence of ratios of $
 B_{(s)} \to \eta^{(\prime)}$ form factor ratios.  Shaded areas show change of the form factors under the variation of $b_2^{\eta^{(\prime)},g} = 0 \pm 20$. Solid line denotes the result for $b_2^{\eta^{(\prime)},g} = 0$, dashed-dotted for $b_2^{\eta^{(\prime)},g} = 20$ and dashed line for $b_2^{\eta^{(\prime)},g} = -20$.}
\label{fig-BgluonR}
%\end{center}
\end{figure}
%}

From Fig.\ref{fig-Dgluon} and Fig.\ref{fig-Bgluon} we see that the gluonic corrections are much larger for $B_{(s)}, D_{(s)} \to \eta^{\prime}$ decays then for $M \to \eta$, as expected. Also the gluonic corrections are larger in $D_{(s)}$ decays. It is obvious that even in ratios of form factors the gluonic contributions give main error and that it would be difficult to constrain $b_2$, unless all $M \to \eta^{(\prime)}$ semileptonic transitions are measured, Fig. 6 and 9. 

We can now investigate $\rm{SU(3)_F}$ approximations from (\ref{eq:SU3}). By using the obtained results and the result for $f_{BK}$ from \cite{DupliJa} we obtain
\ba
|f_{B_s\eta}^{+}|^{\rm calc} = 0.212^{+0.015}_{-0.013}  \quad {\rm vs} \quad |f_{B_s\eta}^{+, \rm approx}| = 0.225^{+0.032}_{-0.026}  \,, \nonumber \\
f_{B_s\eta^{\prime}}^{+, \rm calc} = 0.252^{+0.023}_{-0.020}  \quad {\rm vs} \quad |f_{B_s\eta^{\prime}}^{+, \rm approx}| = 0.278^{+0.038}_{-0.031}  \,. \nonumber \\  
\ea
We can note that the approximation works quite well although somewhat better for 
$M \to \eta$ decays than for $M \to \eta^{\prime}$ transitions. 

There exists LCSR calculations of $f_{D_s\eta}^+$ form factor \cite{ColangeloFazio,Azizi}. In these papers the $f_{D_s\eta^{\prime}}^+$ form factor is then obtained by using the relation
\ba
\frac{f_{D_s\eta^{\prime}}^+}{f_{D_s\eta}^+} = \cot \phi \,. 
\ea 
While their predictions for $f_{D_s\eta}^+$ agree with ours, the use of the above approximative relation which neglects the gluonic contributions gives somewhat 
larger $f_{D_s\eta^{\prime}}^+$ form factor then the one obtained here,  (\ref{eq:results52},\ref{eq:results54}).

There exist also recent lattice results on $D_s \to \eta^{(\prime)}$ form factors \cite{Bali}. These transitions at the lattice are challenging due to the presence
of disconnected quark-line contributions and in \cite{Bali} only the scalar 
$f_{D_s \eta^{(\prime)}}^0$ form factors are calculated, which at $q^2 =0$ are equal to the $f^+$.  By comparing the results one can see that the lattice predictions
give $f_{D_s \eta^{\prime}}^+ < f_{D_s \eta}^+$, which is just opposite in LCSR for all $M_s \to \eta^{(\prime)}$ transitions. 
The tendency  $f_{M \eta^{\prime}}^+ < f_{M \eta}^+$ in LCSR is established for non-strange meson decays, see results in (\ref{eq:results52},\ref{eq:results54}).

\section{Phenomenological applications}

In this section we comment on some phenomenological results for semileptonic 
$D_{(s)} \to \eta^{(\prime)}$  and $B_{(s)} \to \eta^{(\prime)}$ decays which include the calculated from factors.
To be able to calculate the branching ratio we need the form factor extracted in whole accessible kinematical regions. For $D_{(s)}$ decays the LCSR are applicable only in the region $q^2 \ll m_c^2$ and for $B_{(s)}$ the region is $0 < q^2 < 12$ GeV.

The are many parametrization for calculating the shape of form factors at $q^2 \neq 0$. All of them work equally well and therefore we decided to use the most simplest one \cite{BK}:
\ba
f^{{+,T}}_{M\eta^{(\prime)}} (q^2)_{|{\rm fit}} &=& f^{(+,T)}_{M\eta^{(\prime)}}(0)\frac{1}{(1 - q^2/m_{H^{\ast}}^2) (1 - \alpha^{+,T}\, q^2/m_{H^{\ast}}^2 )} \nonumber \\
f^{{0}}_{M\eta^{(\prime)}} (q^2)_{|{\rm fit}} &=& f^{0}_{M\eta^{(\prime)}}(0)\frac{1}{(1 - \alpha^0 \, q^2/m_{H^{\ast}}^2)}
\label{eqddBKform}
\ea
where the extrapolation of the form factors is performed just by fitting one parameter $\alpha^i$ for each of the decays and using the appropriate vector meson resonances $m_H^{\ast}$, Tab.6, while the normalization is given by the form factors at $q^2 = 0$.  
The fitted parameters $\alpha^i$ for $D_{(s)}$ form factors are 
%\ba
%\alpha^+_{D\eta} &=& 0.165 \pm 0.006 \,, \quad \alpha^+_{D\eta'} = 0.19 \pm 0.05 
%\nonumber \\
%\alpha^+_{D_s\eta} &=& 0.198 \pm 0.005 \,, \quad  \alpha^+_{D_s\eta'} = 0.20 \pm 0.03
%\ea
\ba
\alpha^+_{D\eta} &=& 0.17 \pm 0.07 \,, \quad \alpha^+_{D\eta'} = 0.20 \pm 0.20 
\nonumber \\
\alpha^+_{D_s\eta} &=& 0.20 \pm 0.01 \,, \quad  \alpha^+_{D_s\eta'} = 0.20 \pm 0.04
\ea
while for $B_{(s)}$ are as follows:
%\ba
%\alpha^+_{B\eta} &=& 0.462 \pm 0.002 \,, \quad \alpha^0_{B\eta} = 1.00 \pm 0.01 
%\,, \quad  \alpha^T_{B\eta} = 0.494 \pm 0.005 \nonumber \\
%\alpha^+_{B\eta'} &=& 0.45 \pm 0.02 \,, \quad \alpha^0_{B\eta'} = 1.00 \pm 0.09 
%\,, \quad  \alpha^T_{B\eta'} = 0.47 \pm 0.04 \nonumber \\
%\alpha^+_{B_s\eta} &=& 0.505 \pm 0.003 \,, \quad \alpha^0_{B_s\eta} = 1.08 \pm 0.01 
%\,, \quad  \alpha^T_{B_s\eta} = 0.193 \pm 0.002 \nonumber \\
%\alpha^+_{B_s\eta'} &=& 0.433 \pm 0.003 \,, \quad \alpha^0_{B_s\eta'} = 1.09 \pm 0.05 
%\,, \quad  \alpha^T_{B_s\eta'} = 0.51 \pm 0.02 \,.
%\ea
\ba
\alpha^+_{B\eta} &=& 0.46 \pm 0.06 \,, \quad \alpha^0_{B\eta} = 1.00 \pm 0.28 
\,, \quad  \alpha^T_{B\eta} = 0.49 \pm 0.05 \nonumber \\
\alpha^+_{B\eta'} &=& 0.45 \pm 0.17 \,, \quad \alpha^0_{B\eta'} = 1.00 \pm 0.35 
\,, \quad  \alpha^T_{B\eta'} = 0.48 \pm 0.11 \nonumber \\
\alpha^+_{B_s\eta} &=& 0.51 \pm 0.02 \,, \quad \alpha^0_{B_s\eta} = 1.08 \pm 0.06 
\,, \quad  \alpha^T_{B_s\eta} = 0.19 \pm 0.11 \nonumber \\
\alpha^+_{B_s\eta'} &=& 0.49 \pm 0.03 \,, \quad \alpha^0_{B_s\eta'} = 1.09 \pm 0.11 
\,, \quad  \alpha^T_{B_s\eta'} = 0.52 \pm 0.04 \,.
\ea
The large uncertinties of the $\alpha$ parameters are predominantly coming from the large uncertinties in the $\eta - \eta'$ mixing parameters. 

The semileptonic $D_{(s)} \to \eta^{(\prime)} e \nu_e$ and $B \to \eta^{(\prime)} e \nu_e$ decay rates are calculated by
\ba
\Gamma(H \to \eta^{(\prime)}l \bar{\nu_l}) = \frac{G_F^2 |V_{Qq}|^2}{192 \pi^3  m_{H}^3} \int_{m_l^2}^{(m_{H} - m_{\eta^{(\prime)}})^2} dq^2 
\lambda^{3/2}(q^2) |f^+_{H\eta^{(\prime)}}(q^2)|^2 \,,
\ea
where $H = D,D_s,B$ and $\lambda(q^2) 
= 
(m_H^2+ m_{\eta^{(\prime)}}^2 - q^2 )^2 - 4 m_H^2 m_{\eta^{(\prime)}}^2$ and  
$V_{Qq} = V_{cd},V_{cs},V_{ub}$ depending if $D^+,D_s^+$ or $B^+$ meson is decaying, respectively.
Values for the CKM matrix elements are taken from \cite{PDG2014}: 
$
V_{cd}  =  0.225, 
V_{cs}  =  0.973, 
V_{ub}  =  0.0035, 
V_{ts} = 0.0405,
V_{tb}  = 0.999 .
$
(for $V_{ub}$ we used newly determined average value also from \cite{PDG2014}). 

For the rare $B_s \to \eta^{(\prime)} l^+ l^- (\nu \bar{\nu})$ decays we use the effective Standard Model hamiltonian for $b \to s l^+ l^- (\nu \bar{\nu})$ transitions \cite{Buchalla} and calculate decay rates as \cite{DeFazio1}
\ba
\Gamma(B_s \to \eta^{(\prime)}l^+l^-) = \frac{G_F^2 |V_{tb}V_{ts}^{\ast}|^2 \alpha^2}{512 \pi^5 m_{B_s}^3} \int_{4 m_l^2}^{(m_{B_s} - m_{\eta^{(\prime)}})^2} dq^2 
\lambda^{1/2}(q^2) \sqrt{1 - \frac{4 m_l^2}{q^2}} \frac{1}{3 q^2} 
I_{\eta^{\prime)}}(q^2) 
\nonumber \\
\ea
where 
\ba
I_{\eta^{(\prime)}}(q^2) &=& 6 m_l^2 (m_{B_s}^2 - m_{\eta^{(\prime)}}^2)^2 |C_{10} (\mu) f^0_{B_s\eta^{(\prime)}}(q^2)|^2
+ (q^2 - 4 m_l^2) \lambda(q^2)|C_{10} (\mu) f^+_{B_s\eta^{(\prime)}}(q^2)|^2  
\nonumber \\
&& + (q^2 + 2 m_l^2) \lambda(q^2) |C_9 (\mu) f^+_{B_s\eta^{(\prime)}}(q^2) - 2 \frac{m_b + m_s}{m_{B_s} + m_{\eta^{(\prime)}}} C_7(\mu) f^T_{B_s\eta^{(\prime)}}(q^2)|^2 \,, \nonumber 
\ea
and 
\ba
\Gamma(B_s \to \eta^{(\prime)}\nu\bar{\nu}) = 3 \frac{|C_L|^2}{96 \pi^3 m_{B_s}^3} \int_{0}^{(m_{B_s} - m_{\eta^{(\prime)}})^2} dq^2 
\lambda^{3/2}(q^2) |f^+_{B_s\eta^{(\prime)}}(q^2)|^2 \,,
\ea
where $C_L = G_F/\sqrt{2} \alpha/(2 \pi \sin^2\theta_W) V_{tb} V_{ts}^{\ast} \eta_X X(x_t)$ \cite{DeFazio1}. 
For the Wilson coefficients we use the following values
\ba
C_7 = - 0.3031 \,, C_9 = 4.1696 \,, C_{10} = - 4.4641 \,, C_L = 2.74 \cdot 10^{-9}
\,.
\ea

Our predicted branching ratios for various $M \to \eta^{(\prime)}$ decays are given in Tab.1.
By comparing with the existing calculations performed in different models \cite{DeFazio1,Azizi2,Choi,Faustov} we agree quite well, expect that we predict somewhat larger branching ratios for $B_s \to \eta^{(\prime)} \tau^+ \tau^-$ decays. 

Because of the larger errors in $B,D \to \eta^{(\prime)}$ decays, $M_s \to \eta^{(\prime)}$ would be better for extraction of the unknown $b_2^{\eta^{(\prime)},g}$ parameter, but measurements of $M_s$ decays still have to achieve sufficient 
precision, in particular $Br(B_s \to \eta^{\prime} l^+l^-)$ and
$Br(B_s \to \eta l^+l^-)$ are challenging with the branching ratio of $O(10^{-7} - 10^{-8})$ but they could be measured at future SuperB and SuperKEK experiments.
%%%%%%%%%%%%%%%%%%%%%%%%%%%%%%%%%%%%%%%%%%%
%\TABLE[h]{
\begin{table}[h]
\begin{center}
\begin{tabular}{|c|c|c|}
\hline
Branching ratio &
Predicted value &
Experiment \\
\hline
${\rm Br} \left( D^+ \to \eta\, e^+ \nu_e \right)$ &  $(14.24\pm 10.98)\cdot 10^{-4}$ &  $(11.4 \pm 0.9 \pm 0.4) \cdot 10^{-4}$ \protect{\cite{exp1}}\\
${\rm Br} \left( D^+ \to \eta' e^+ \nu_e \right)$ & $(1.52\pm 1.17)\cdot 10^{-4} $ 
&  $(2.16 \pm 0.53 \pm 0.07) \cdot 10^{-4}$ \protect{\cite{exp1}}\\
\hline
$\frac{\displaystyle \Gamma \left( D^+ \to \eta' e^+ \nu_e \right)}{\displaystyle \Gamma \left( D^+ \to \eta\, e^+ \nu_e \right)}$  &  $0.10 \pm  0.11$  &  $0.19\pm 0.09$ \protect{\cite{exp2}}\\
\hline
${\rm Br} \left( D^+_s \to \eta\, e^+ \nu_e \right)$ & $(2.40\pm 0.28)\%$ 
& $(2.48 \pm 0.29)\%$ \protect{\cite{exp2}} \\
${\rm Br} \left( D^+_s \to \eta' e^+ \nu_e \right)$ &  $(0.79\pm 0.14)\%$ 
& $(0.91 \pm 0.33)\%$ \protect{\cite{exp2}}  \\
\hline
%$\frac{\displaystyle \Gamma \left( D^+ \to \eta' e^+ \nu_e \right)}{\displaystyle \Gamma \left( D^+ \to \eta\, e^+ \nu_e \right)}$  &  $0.10 \pm  0.11$  &  $0.19\pm 0.09$ \protect{\cite{exp2}}\\
$\frac{\displaystyle \Gamma \left( D^+_s \to \eta' e^+ \nu_e \right)}{\displaystyle \Gamma\left( D^+_s \to \eta\, e^+ \nu_e \right)}$ & $ 0.33 \pm 0.07$ & $0.36 \pm 0.14$\protect{\cite{exp1}} \\
\hline
${\rm Br} \left( B^+ \to \eta\, e^+ \nu_e \right)$ & $ (0.44\pm 0.25)\cdot 10^{-4}$ 
& $(0.44 \pm 0.23 \pm 0.11)\cdot 10^{-4}$ \protect{\cite{exp3}}  \\
& &  $(0.36 \pm 0.05 \pm 0.04)\cdot 10^{-4}$ \protect{\cite{exp4}} \\
 \hline
${\rm Br} \left( B^+ \to \eta' e^+ \nu_e \right)$ & $ (0.19\pm 0.11)\cdot 10^{-4}$  
& $(2.66 \pm 0.80 \pm 0.56)\cdot 10^{-4}$ \protect{\cite{exp3}}  \\
& &  $(0.24 \pm 0.08 \pm 0.03)\cdot 10^{-4}$ \protect{\cite{exp4}}  \\
\hline
$\frac{\displaystyle \Gamma \left( B^+ \to \eta' e^+ \nu_e \right)}{\displaystyle \Gamma \left( B^+ \to \eta\, e^+ \nu_e \right)}$ &  $0.43 \pm 0.34$  & $0.67 \pm 0.24 \pm 0.1$ \protect{\cite{exp4}} \\ 
\hline
${\rm Br} \left( B_s \to \eta\, l^+ l^- \right)_{l = e,\mu}$ 
& $(2.80 \pm 0.36)\cdot 10^{-7}$   &   \\
${\rm Br} \left( B_s \to \eta\, \tau^+ \tau^- \right)$ 
& $(1.53 \pm 0.18)\cdot 10^{-7}$  &   \\
\hline 
${\rm Br} \left( B_s \to \eta' l^+ l^- \right)_{l = e,\mu}$ 
& $(2.85 \pm 0.48)\cdot 10^{-7}$  & \\
${\rm Br} \left( B_s \to \eta' \tau^+ \tau^- \right)$ 
& $(0.75 \pm 0.14)\cdot 10^{-7}$  & \\
\hline
${\rm Br} \left( B_s \to \eta \, \nu \bar{\nu} \right)$ 
& $(20.5 \pm 2.8)\cdot 10^{-7}$ &   \\
${\rm Br} \left( B_s \to \eta' \, \nu  \bar{\nu} \right)$ 
& $(14.8 \pm 2.0)\cdot 10^{-7}$ &\\
\hline
\end{tabular}
\end{center}
\caption{\it Predicted branching fractions of various $D_{(s)}, B_{(s)} \to \eta^{(\prime)}$ semileptonic decays.}
\label{tab-predictions}
\end{table}
%}
%%%%%%%%%%%%%%%%%%%%%%%%%%%%%%%%%%%%%%%%%%%%%%%%%%
%%%%%%%%%%%%%%%%%%%%%%%%%%%%%%%%%%%%%%%%%%%%%%%%%%%%%%%%%%%%%%%%%%%%%%%%%%%%%%%%%%%%%%%
\section{Summary}

We have investigated $B, B_{s}\to \eta^{(\prime)}$ and $D, D_{s}\to \eta^{(\prime)}$ form factors ($f^+, f^0$ and $f^T$) by including $m_{\eta^{(\prime)}}^2$ corrections 
in the leading (up to the twist-four) and next-to-leading order (up to the twist-three) in QCD, as well as  gluonic contributions to the form factors at the leading twist 
in the framework of the QCD light-cone sum rules and have also taken SU(3)-flavour breaking corrections and the axial anomaly contributions to the distribution amplitudes 
consistently into account. The two-gluon twist-2 contributions are calculated for all $f^+$, $f^0$ and $f^T$ form factors.

We have given the values and shapes at $q^2 \neq 0$ of all calculated form factors and have shown predicted ratios for some semileptonic $B, B_{s}\to \eta^{(\prime)}$ 
and $D, D_{s}\to \eta^{(\prime)}$ decay modes. 
 
With the determined form factors of transitions $B,B_s \to \eta^{(\prime)}$ it will be possible to analyze consistently nonleptonic decays to charmonia and to test the 
factorization hypothesis in such transitions which we be a subject of the future investigations.   
%
%
%%%%%%%%%%%%%%%%%%%%%%%%%%%%%%%%%%%%%%%%%%%%%%%%%%%%%%%%%%%%%%%%%%%%%%%%%%%%%%%%
\section*{Acknowledgments}
We are grateful to D. Be\v cirevi\'c and K. Passek-Kumeri\v cki 
for useful discussions. 
The work is supported by the Croatian Science Foundation (HrZZ) project "Physics of Standard Model and Beyond", HrZZ 5169 and by the Munich Institute for Astro- and Particle Physics (MIAPP) of the DFG cluster of excellence "Origin and Structure of the Universe".
%%%%%%%%%%%%%%%%%%%%%%%%%%%%%%%%%
%\newpage
%%%%%%%%%%%%%%%%%%%%%%%%%%%%%%%%%%%%%%%%%%%%%%%%%%%%%%%%%%%%%%%%%%%%%%%%%%%%%%%
\appendix

\section{Explicit results for $f^+$, $f^0$ and $f^T$ form factors at the leading order in $B, B_{s}\to \eta^{(\prime)}$ and $D, D_{s}\to \eta^{(\prime)}$ transitions}
 
The  leading ${\cal O}(\alpha_s)$ part of the $f_{B_{(s)}\eta^{(\prime)}}^+$ LCSR, (\ref{eq:fplusLCSR}), has the following expression ($P = \eta,\eta^\prime$; $r=q$ for $B_q \to P$ and $r=s$ for $B_s \to P$; for $D,D_s$ the same expressions are valid with the replacement $m_b \to m_c$):
\ba
&& F_{0, B_r \to P}(q^2,M^2,s_0^B)= m_b^2 \int\limits_{u_0}^1 du\,
e^{-\frac{m_b^2-q^2\bar{u} + m_P^2 u \bar{u}}{u M^2}} 
\Bigg\{
\frac{ F_P^{(r)} \varphi_{2P}^{(r)}(u)}{u}\qquad\qquad\qquad\qquad
\nonumber
\\
&& \qquad +
\frac{1}{2 m_r m_b}
\Bigg[ \phi_{3P}^{(r)p}(u)
+\frac{1}{6}\Bigg( 2 \frac{\phi_{3P}^{(r)\sigma}(u)}{u}
-\frac{1}{m_b^2 -q^2 + u^2 m_P^2} 
\bigg ( ( m_b^2 +q^2 - u^2 m_P^2 ) \frac{d \phi_{3P}^{(r)\sigma}(u)}{du} 
\nonumber \\
&& \qquad \quad - \frac{4 u m_P^2 m_b^2}{m_b^2 -q^2 + u^2 m_P^2}  \phi_{3P}^{(r)\sigma}(u) \bigg )\Bigg)\Bigg]
\nonumber 
\\
&& \qquad +\frac{F_P^{(r)}}{m_b^2-q^2 + u^2 m_P^2}
\Bigg[ 
u\overline{\psi}_{4P}^{(r)}(u)+ \left ( 1 - \frac{2 u^2 m_P^2}{m_b^2-q^2 + u^2 m_P^2} \right )\int\limits_0^u dv \overline{\psi}_{4P}^{(r)}(v)
\nonumber 
\\
&& \qquad  
- \frac{m_b^2}{4}\frac{u}{m_b^2-q^2 + u^2 m_P^2} 
\left ( \frac{d^2}{du^2} - \frac{6 u m_P^2}{m_b^2-q^2 + u^2 m_P^2} \frac{d}{du} + 
\frac{12 u m_P^4}{(m_b^2-q^2 + u^2 m_P^2)^2} \right ) \overline{\phi}_{4P}^{(r)}(u) 
\nonumber 
\\
&& \qquad - \left ( \frac{d}{du} - \frac{2 u m_P^2}{m_b^2-q^2 + u^2 m_P^2} \right ) 
\left ( \left (\frac{F_{3P}^{(r)}}{m_b F_P^{(r)}} \right ) I_{3r}(u) + I_{4P}^{(r)}(u)  \right )
\nonumber \\
&& \qquad
-\frac{ 2 u m_P^2 }{m_b^2-q^2 + u^2 m_P^2} 
\left ( u\frac{d}{du} + \left (1- \frac{4 u^2 m_P^2}{m_b^2-q^2 + u^2 m_P^2} \right ) \right ) \overline{I}_{4P}^{(r)}(u) 
\nonumber 
\\
&& \qquad + \frac{ 2 u m_P^2 (m_b^2-q^2 - u^2 m_P^2)}{(m_b^2-q^2 + u^2 m_P^2)^2} 
\left ( \frac{d}{du} - \frac{6 u m_P^2}{m_b^2-q^2 + u^2 m_P^2} \right ) \int_u^1 d\xi \overline{I}_{4P}^{(r)}(\xi)
\Bigg]
\Bigg\}
\nonumber \\
&& \qquad +  \frac{m_b^4 F_P^{(r)} e^{-\frac{m_b^2}{ M^2}}}{4 (m_b^2 -q^2 +  m_P^2)^2}
\bigg ( \frac{d \overline{\phi}_{4P}^{(r)}(u)}{du} \bigg )_{u \to 1}  \, , 
\label{eq:fplusBpiLCSRcontrib}
\ea
where $\bar{u}=1-u$, $u_0=\left (q^2 - s_0^B + m_P^2 + \sqrt{(q^2 - s_0^B + m_P^2)^2 - 4 m_P^2 (q^2 - m_b^2)} \right )/(2 m_P^2) $, $F_\eta^{(q)} =\cos\phi f_q/\sqrt{2},F_{\eta^\prime}^{(q)} =\sin\phi f_q/\sqrt{2}$, $F_{\eta}^{(s)} = - \sin\phi f_s$, $F_{\eta^\prime}^{(s)} = \cos\phi f_s$, and similarly for the two-particle twist-three DAs $\phi_{3P}^{p,\sigma}$: $\phi_{3\eta}^{(q) p,\sigma} =\cos\phi \, \phi_{3q}^{p,\sigma}/\sqrt{2} ,\phi_{3\eta^\prime}^{(q)p,\sigma} =\sin\phi \,\phi_{3q}^{p,\sigma}/\sqrt{2}$, $\phi_{3\eta}^{(s)p,\sigma} = - \sin\phi \,\phi_{3s}^{p,\sigma}$, $\phi_{3\eta^\prime}^{(s)p,\sigma} = \cos\phi \,\phi_{3s}^{p,\sigma}$. 
Also, 
\ba
\overline{\psi}_{4P}^{(r)}(u) = \psi_{4P}^{(r){\rm tw}}(u) + \frac{h_r}{f_r} \psi_{4P}^{(r){\rm mass}}(u) \,, \nonumber \\
\overline{\phi}_{4P}^{(r)}(u) = \phi_{4P}^{(r){\rm tw}}(u) + \frac{h_r}{f_r} \phi_{4P}^{(r){\rm mass}}(u) \,.
\ea

In the case of the twist-2 DA, we will express the decay constants $F_P^{(q)}$ in the SO basis and take the different evolution of $f_P^{(1)}$ and $f_P^{(8)}$ into account:
\begin{eqnarray}
f_P^{(1)}(\mu) = f_P^{(1)}(\mu_0) \, , \nonumber \\
f_P^{(8)}(\mu) = f_P^{(8)}(\mu_0)\left [ 1 + \frac{2 n_f}{\pi \beta_0} \left ( 
\alpha_s(\mu) - \alpha_s(\mu_0) \right ) \right ] \, , \nonumber \\
\end{eqnarray} 
at $m_0 = 1$ GeV, the energy at which the FKS parameters are determined and 
\begin{eqnarray}
F_P^{(q)} \phi_{2 P}^{(q)} = C_P^q\frac{1}{\sqrt{3}} \left ( \sqrt{2} f^1 + f^8  \right ) \phi_{2P} \,,\nonumber \\
F_P^{(s)} \phi_{2 P}^{(q)} = C_P^s\frac{1}{\sqrt{3}} \left ( f^1 - \sqrt{2} f^1 + f^8 \right ) \phi_{2P} \,,\nonumber \\
\end{eqnarray} 
with $C_\eta^q = \cos\phi/\sqrt{2}$, $C_{\eta^\prime}^q = \sin\phi/\sqrt{2}$, $C_\eta^s = -\sin\phi$, $C_{\eta^\prime}^s = \cos\phi$ and 
\begin{eqnarray}
\phi_{2\eta} = \phi_{2\eta^{\prime}} = 6 u (1-u) \left (1 + \sum_{i=2,4} a_i C_i^{3/2}(2u-1) \right )\,.
\end{eqnarray}
Numerically, 
\begin{eqnarray}
f_1(\mu_0) = (1.17 \pm 0.03) f_\pi \,, \nonumber \\
f_8(\mu_0) = (1.26 \pm 0.04) f_\pi \,. \nonumber
\end{eqnarray}
The 
short-hand notations introduced for the integrals over three-particle DA's are
\footnote{In the paper \cite{DupliJa}, dealing with $B_{(s)} \to K$ form factors, in eq.(A.2) there was a misprint in the function $I_{3K}(u)$, the factor of 3 was missing.
The correct expression has the same form as $I_{3r}(u)$ given here in (\ref{eq:fplusBpiLCSR3part}).}:
\ba
&& I_{3r}(u)=\int\limits_0^u \!d\alpha_1\!\!\!
\int\limits_{(u-\alpha_1)/(1-\alpha_1)}^1\!\!\!\!\! \frac{dv}{v} \,\,
\left [ 4 v p \cdot q - 3 (1 - 2v) u m_P^2 \right ] \Phi_{3r}(\alpha_i)
\Bigg|_{\begin{array}{l}
\alpha_2=1-\alpha_1-\alpha_3,\\
\alpha_3=(u-\alpha_1)/v
\end{array}
}
\,,
\nonumber
\\
&& I_{4P}^{(r)}(u)=\int\limits_0^u\! d\alpha_1\!\!\!
\int\limits_{(u-\alpha_1)/(1-\alpha_1)}^1\!\!\!\!\! \frac{dv}{v} \,\,
\Bigg[ 2 \Psi_{4P}^{(r)}(\alpha_i)-  \Phi_{4P}^{(r)}(\alpha_i) + 2 \widetilde{\Psi}_{4P}^{(r)}(\alpha_i) - \widetilde{\Phi}_{4P}^{(r)}(\alpha_i) 
\Bigg]
\Bigg|_{\begin{array}{l}
\alpha_2=1-\alpha_1-\alpha_3,\\
\alpha_3=(u-\alpha_1)/v
\end{array}
}
\,.
\nonumber
\\
&& \overline{I}_{4P}^{(r)}(u)=\int\limits_0^u\! d\alpha_1\!\!\!
\int\limits_{(u-\alpha_1)/(1-\alpha_1)}^1\!\!\!\!\! \frac{dv}{v} \,\,
\Bigg[
\Psi_{4P}^{(r)}(\alpha_i) + \Phi_{4P}^{(r)}(\alpha_i) + 
 \widetilde{\Psi}_{4P}^{(r)}(\alpha_i)+ \widetilde{\Phi}_{4P}^{(r)}(\alpha_i) 
\Bigg]
\Bigg|_{\begin{array}{l}
\alpha_2=1-\alpha_1-\alpha_3,\\
\alpha_3=(u-\alpha_1)/v
\end{array}
}
\,.
\label{eq:fplusBpiLCSR3part}
\ea

The leading order LCSR for $f^+_{BK} + f^-_{BK}$, (\ref{eq:fplminLCSR}), has the form 
\ba
&& \widetilde{F}_{0,B_r \to P}(q^2,M^2,s_0^B)= 
m_b^2 \int\limits_{u_0}^1 du\, e^{-\frac{m_b^2-q^2\bar{u} + m_P^2 u \bar{u}}{u M^2}}
\Bigg\{\frac{1}{2 m_r m_b}\Bigg(\frac{\phi_{3P}^{(r)p}(u)}{u}
+\frac{1}{6u}\frac{d\phi_{3P}^{(r)\sigma}(u)}{du}\Bigg)
\nonumber
\\
&& \qquad +\frac{F_P^{(r)}}{m_b^2-q^2 + u^2 m_P^2} 
\Bigg [ \overline{\psi}_{4P}^{(r)}(u) - \frac{2 u m_P^2}{m_b^2 - q^2 + u^2 m_P^2} \int_0^u dv \overline{\psi}_{4P}^{(r)}(v) 
\nonumber \\
&& \qquad + m_P^2 \left (\frac{d}{du} - \frac{2 u m_P^2}{m_b^2 - q^2 + u^2 m_P^2} \right ) 
\left (\frac{F_{3P}^{(r)}}{m_b F_P^{(r)}} \right ) \widetilde{I}_{3r}(u) 
\nonumber \\
&& \qquad + \frac{2 u m_P^2}{m_b^2 - q^2 + u^2 m_P^2} \bigg ( \frac{d^2}{du^2} - \frac{6 u m_P^2}{m_b^2 - q^2 + u^2 m_P^2} \frac{d}{du} 
+ \frac{12 u^2 m_P^4}{(m_b^2 - q^2 + u^2 m_P^2)^2} \bigg ) \int_u^1 d\xi \overline{I}_{4P}^{(r)}(\xi)
\Bigg ]
\Bigg\}
\,.
\nonumber \\
\label{eq:fplminBpiLCSRcontrib}
\ea
where 
\ba
&& \widetilde{I}_{3r}(u)=\int\limits_0^u \!d\alpha_1\!\!\!
\int\limits_{(u-\alpha_1)/(1-\alpha_1)}^1\!\!\!\!\! \frac{dv}{v} \,\,
\left [ (3 - 2v) \right ] \Phi_{3r}(\alpha_i)
\Bigg|_{\begin{array}{l}
\alpha_2=1-\alpha_1-\alpha_3,\\
\alpha_3=(u-\alpha_1)/v
\end{array}
}
\,.
\ea
Finally, the leading order LCSR for the penguin form factor, (\ref{eq:fTLCSR}), reads: 
\ba
&& F_{0,B_r \to P}^{T}(q^2,M^2,s_0^B)= m_b \int\limits_{u_0}^1 
du\, e^{-\frac{m_b^2-q^2\bar{u} + m_P^2 u \bar{u}}{uM^2}} 
\Bigg\{\frac{F_P^{(r)}\phi_{2P}^{(r)}(u)}{u}
\nonumber \\
&& \quad 
-\frac{m_b}{6 m_r (m_b^2-q^2 + u^2 m_P^2)}
\left (\frac{d\phi_{3P}^{(r)\sigma}(u)}{du} - \frac{2 u m_P^2}{m_b^2 - q^2 + u^2 m_P^2} \phi_{3P}^{(r)\sigma}(u) \right )
\nonumber
\\
&& \quad
+\frac{F_P^{(r)}}{m_b^2-q^2 + u^2 m_P^2}
\Bigg[\left ( \frac{d}{du} - \frac{2 u m_P^2}{m_b^2 - q^2 + u^2 m_P^2} \right )
\bigg ( \frac{1}{4} \overline{\phi}_{4P}^{(r)}(u) - I_{4P}^{(r)T}(u) \bigg )
\nonumber \\
&& \quad 
- \frac{m_b^2\,u}{4(m_b^2-q^2 + u^2 m_P^2)}\bigg (\frac{d^2}{du^2} - \frac{6 u m_P^2}{m_b^2 - q^2 + u^2 m_P^2} 
\frac{d}{du} + \frac{12 u m_P^4}{(m_b^2 - q^2 + u^2 m_P^2)^2} \bigg ) \overline{\phi}_{4P}^{(r)}(u)
\Bigg]\Bigg\}
\nonumber \\
&& +  \frac{m_b^3 F_P^{(r)} e^{-\frac{m_b^2}{ M^2}}}{4 (m_b^2 -q^2 +  m_P^2)^2}
\bigg ( \frac{d \overline{\phi}_{4P}^{(r)}(u)}{du} \bigg )_{u \to 1}
\, 
\label{eq:fTBpiLCSRcontrib}
\ea
and
\ba
&I_{4P}^{(r)T}(u)&= \int\limits_0^u\! d\alpha_1\!\!\!
\int\limits_{(u-\alpha_1)/(1-\alpha_1)}^1\!\!\!\!\! \frac{dv}{v} \,\,
\Bigg[2\Psi_{4P}^{(r)}(\alpha_i)-(1-2v)\Phi_{4P}^{(r)}(\alpha_i)
\nonumber
\\
&&  \qquad\qquad\qquad 
+ 2(1-2v)\widetilde{\Psi}_{4P}^{(r)}(\alpha_i)-
\widetilde{\Phi}_{4P}^{(r)}(\alpha_i)\Bigg]
\Bigg|_{\begin{array}{l}
\alpha_2=1-\alpha_1-\alpha_3,\\
\alpha_3=(u-\alpha_1)/v
\end{array}
}\,.
\label{eq:fTBpiLCSR3part}
\ea
%%%%%%%%%%%%%%%%%%%%
%Note the appearance of the surface terms in form factors above. 

The expressions for $f_{D_{(s)}}^{+,0,T}$ from factors follows from above, by replacing $m_b$ by $m_c$. 

%%%%%%%%%%%%%%%%%%%%%%%%%%%%%%%%%%%%%%%%%%%%%%%%%%%%%%%%%%%%%%%%%%%%%%%%%%%%%%%%%

\section{Parameters used in the calculation}

In this appendix we summarize the parameters used in the calculation of $f_{M\eta^{(\prime)}}$ form factors as well as in the calculation of 
$f_M$ decay constants, Tables 2-5.  
In Table 6. we summarize meson masses, lifetimes and vector resonances used in the calculation of phenomenological predictions for semileptonic $M \to \eta^{(\prime)}$ decays. 
%%%%%%%%%%%%%%%%%%%%%%%%%%%%%%%%%%%%%%%%%%%
%\TABLE[h]{
\begin{table}[h]
\begin{center}
\begin{tabular}{|c|c|}
\hline
Parameter & Value at $\mu=1$ GeV \\
\hline
$a_2^\pi$  &  $0.17 \pm 0.08$ \protect{\cite{KMOW}}\\
$a_4^\pi$ & $0.06 \pm 0.10$ \protect{\cite{KMOW}}\\
$a_{> 4}^\pi$ & 0 \\
\hline
$f_{3\pi}$ & $0.0045\pm 0.0015$ GeV$^2$ \\
$\omega_{3\pi}$ & $-1.5\pm 0.7$ \\
%$\lambda_{3K}$ & $1.6 \pm 0.4$ \\
\hline
$\delta^2_{\pi}$& $0.18\pm 0.06$  GeV$^2$ \\
%$\epsilon_{\pi}$ & $\frac{21}{8} (0.2 \pm 0.1)$ \\
$\omega_{4\pi}$ & $0.2\pm 0.1$ \\
%$\kappa_{4K}$ & $-0.09 \pm 0.02$ \\
\hline
\hline
$f_{3K}$ & $0.0045\pm 0.0015$ GeV$^2$ \\
$\omega_{3K}$ & $-1.2\pm 0.7$ \\
$\lambda_{3K}$ & $1.6 \pm 0.4$ \\
\hline
$\delta^2_{K}$& $0.20\pm 0.06$  GeV$^2$ \\
$\omega_{4K}$ & $0.2\pm 0.1$ \\
$\kappa_{4K}$ & $-0.09 \pm 0.02$ \\
\hline
\end{tabular}
\end{center}
\caption{\it Input parameters in DA's \protect{\cite{BBL,DKMMO}}.}
\label{tab-1}
\end{table}
%%%%%%%%%%%%%%%%%%%%%%%%%%%%%%%%%%%%%%%%%%%%%%%%%%
%%%%%%%%%%%%%%%%%%%%%%%%%%%%%%%%%%%%%%%%%%%
%%%%%%%%%%%%%%%%%%%%%%%%%%%%%%%%%%%%%%%%%%
%\TABLE[t]{
\begin{table}
\begin{center}
\begin{tabular}{|c|c|}
\hline
Parameter & Value  \\
\hline
$\overline{m}_b(\overline{m}_b)$ & $4.18 \pm 0.03$ GeV \\
$\overline{m}_c(\overline{m}_c)$ & $1.275 \pm 0.025$ GeV \\
$\overline{m}_u$(2 \,GeV) & $2.3^{+ 0.7}_{- 0.5}$ MeV \\
$\overline{m}_d$(2 \,GeV) & $4.8^{+ 0.5}_{- 0.3}$ MeV \\
$\overline{m}$(2 \,GeV) = $\frac{m_u + m_d}{2}$ & $3.5^{+ 0.7}_{- 0.2}$ MeV \\
$\overline{m}_s$(2 \,GeV)& $95 \pm 5$ MeV \\
%\vspace*{0.2cm}
\hline
$\langle \bar{q}q \rangle $(1 GeV) & $-(246^{+18}_{-19}\,{\rm MeV})^3$ \\
$\langle \bar{s}s \rangle $/$\langle \bar{q}q \rangle $ & $0.8 \pm 0.3$ \\
$\langle \alpha_s/\pi \,GG \rangle $ & $0.012^{+0.006}_{-0.012}\, {\rm GeV^4}$ \\
$m_0^2$ & $0.8 \pm 0.2\,{\rm  GeV}^2$
\\
\hline
$\alpha_s (M_z)$ & $0.1176 \pm 0.002$ \\
\hline
\end{tabular}
\\
\end{center}
\caption{\it Quark masses and additional input parameters for the $f_{B_{(s)}}$ and $f_{D_{(s)}}$ sum rules.}
\label{tab-2}
\end{table}
%%%%%%%%%%%%%%%%%%%%%%%%%%%%%%%%%%%%%%%%%%%%%%%%%%
%mDstar = 2.0102;
%mDSstar = 2.1121;
%mBstar = 5.3252;
%mBSstar = 5.4154;
%%%%%%%%%%%%%%%%%%%%%%%%%%%%%%%%%%%%%%%%%%
%%%%%%%%%%%%%%%%%%%%%%%%%%%%%%%%%%%%%%%%%%
%\TABLE[t]{
\begin{table}
\begin{center}
\begin{tabular}{|c|c|c|c|}
\hline
Decay constant & LCSR \cite{KhodjamPivovarov} &  This work & Fitted $M^2$ and $s_0$ ($\rm{GeV}^2$)\\
\hline
$f_{\pi}$ &   & 130.7 & \\
$f_{K}$ &  & 155 & \\
\hline
$f_{D}$   & $201^{ + 12}_{ - 13}$ & $191 \pm 9$ & $M^2 = 2$, $s_0 = 5$\\
$f_{D_{s}}$ & $238^{ + 13}_{ - 23}$ & $219 \pm 7$ & $M^2 = 2$, $s_0 = 6.3$\\
$f_{B}$  & $207^{ +17}_{- 09}$ & $215 \pm 7$ & $M^2 = 5$, $s_0 = 35.6$\\
$f_{B_s}$ &  $242^{+ 17}_{ - 12}$   & $246 \pm 8$ &  $M^2 = 5.1$, $s_0 = 35.5$\\
\hline
$\frac{f_{D_s}}{f_D}$  &  $1.15^{+0.04}_{-0.05}$  &   $1.15 \pm 0.05$ & \\
$\frac{f_{B_s}}{f_B}$  &  $1.17^{+0.04}_{-0.03}$ & $1.16 \pm 0.05$ & \\
\hline
\end{tabular}
\\
\end{center}
\caption{\it Decay constants used in the paper, the values are in MeV. The decay constants of heavy mesons are obtained from the two-point SR at $O(\alpha_s)$ and agree with those from \cite{DupliJa,DKMMO,KhodjamPivovarov}. The quoted errors are coming only from the variation of $s_0$ and the Borel parameter $M$, since other errors will cancel in the calculation of the form factors. For comparison the recent more complete LCSR results from \cite{KhodjamPivovarov} are given.}
\label{tab-4}
\end{table}
%%%%%%%%%%%%%%%%%%%%%%%%%%%%%%%%%%%%%%%%%%%%%%%%%
%\ba
%M_{B}^2 &=& 18 \pm 2\,{\rm GeV}^2, \quad s_0^{B} = 37 \pm 0.5 \,{\rm GeV}^2, %\nonumber \\ 
%M_{B_s\eta}^2 &=& 17 \pm 1\,{\rm GeV}^2 \quad s_0^{B_s\eta} = 37.5 \pm 0.5\,{\rm %GeV}^2 \nonumber \\
%M_{B_s\eta^{\prime}}^2 &=& 18 \pm 2\,{\rm GeV}^2 \quad s_0^{B_s\eta^{\prime}} = %38 \pm 0.5\,{\rm GeV}^2 \nonumber \\
%M_{D\eta}^2 &=& 5.2 \pm 0.8\,{\rm GeV}^2, \quad s_0^{D\eta} = 7 \pm 0.2\,{\rm %GeV}^2, \nonumber \\ 
%M_{D\eta^{\prime}}^2 &=& 5 \pm 1\,{\rm GeV}^2, \quad s_0^{D\eta^{\prime}} = 5.5 %\pm 0.3 \,{\rm GeV}^2, \nonumber \\ 
%M_{D_s\eta}^2 &=& 8 \pm 0.2\,{\rm GeV}^2 \quad s_0^{D_s\eta} = 7.8 \pm 0.2\,{\rm %GeV}^2 \nonumber \\
%M_{D_s\eta^{\prime}}^2 &=& 6 \pm 1\,{\rm GeV}^2 \quad s_0^{D_s\eta^{\prime}} = %7.5 \pm 0.5 \,{\rm GeV}^2 
%\ea 
%
%%%%%%%%%%%%%%%%%%%%%%%%%%%%%%%%%%%%%%%%%%
%\TABLE[t]{
\begin{table}
\begin{center}
\begin{tabular}{|c|c|}
\hline
Transition & Fitted $M^2$ and $s_0$ parameters of LCSR for  $f^{+,0,T}$\\
\hline
$B \to \eta^{(\prime)}$ & $M_{B}^2 = 18 \pm 2\,{\rm GeV}^2,  s_0^{B} = 37 \pm 0.5 \,{\rm GeV}^2, $\\ 
$B_s \to \eta$ & $M_{B_s\eta}^2 = 17 \pm 1\,{\rm GeV}^2, s_0^{B_s\eta} = 37.5 \pm 0.5\,{\rm GeV}^2$ \\
$B_s \to \eta^{\prime}$ & $M_{B_s\eta^{\prime}}^2 = 18 \pm 2\,{\rm GeV}^2, s_0^{B_s\eta^{\prime}} = 38 \pm 0.5\,{\rm GeV}^2$ \\
$D \to \eta$ & $M_{D\eta}^2 = 5.2 \pm 0.8\,{\rm GeV}^2,  s_0^{D\eta} = 7 \pm 0.2\,{\rm GeV}^2$ \\ 
$D \to \eta^{\prime}$ & $M_{D\eta^{\prime}}^2 = 5 \pm 1\,{\rm GeV}^2 s_0^{D\eta^{\prime}} = 5.5 \pm 0.3 \,{\rm GeV}^2$ \\ 
$D_s \to \eta$ & $M_{D_s\eta}^2 = 8 \pm 0.2\,{\rm GeV}^2 , s_0^{D_s\eta} = 7.8 \pm 0.2\,{\rm GeV}^2 $\\
$D_s \to \eta^{\prime}$ & $M_{D_s\eta^{\prime}}^2 = 6 \pm 1\,{\rm GeV}^2 , s_0^{D_s\eta^{\prime}} = 7.5 \pm 0.5 \,{\rm GeV}^2 $\\
\hline
\end{tabular}
\\
\end{center}
\caption{\it Fitted Borel parameters $M^2$ and the continuum thresholds $s_0$ for each of the decays used to obtain the predicted form factors in the text. }
\label{tab-5}
\end{table}
%%%%%%%%%%%%%%%%%%%%%%%%%%%%%%%%%%%%%%%%%%%%%%%%%
%\TABLE[t]{
\begin{table}
\begin{center}
\begin{tabular}{|c|c||c|c||c|c|}
\hline
Mass & Value (GeV) & Resonance & Mass value (GeV) & Lifetime & Value (ps)\\
\hline
$m_{B^+}$ & 5.2792 & $m_{B^{\ast}}(1^-)$ & 5.3252 & $\tau_{B^+}$ & $1.638 \pm 0.004$ \\
$m_{B_{s}}$ & 5.3667 & $m_{B_s^{\ast}}(1^-)$ & 5.4154 & $\tau_{B_{s}}$ & $1.512 \pm 0.007$ \\
$m_{D^+}$ & 1.8696 & $m_{D^{\ast}}(1^-)$ & 2.0102 & $\tau_{D^+}$ & $1.040 \pm 0.007$ \\
$m_{D_{s}^+}$ & 1.9685 & $m_{D_s^{\ast}}(1^-)$ & 2.1121 & $\tau_{D_{s}^+}$ & $0.500 \pm 0.007$\\
\hline
$m_{\pi^0}$ & 0.1359  & & & & \\
$m_{K^0}$ & 0.4976 & & & & \\
$m_{\eta}$ & 0.5478 & & & &  \\
$m_{\eta^{\prime}}$ & 0.9577 & & & &\\
\hline
\end{tabular}
%\hspace*{2cm}
%\begin{tabular}{|c|c|}
%\hline
%Lifetime & Value (ps)  \\
%\hline
%\tau(D^+) & = & \left(1.040 \pm 0.007 \right)\, {\rm ps} \nonumber \\
%\tau(D^+_s) & = & \left(0.500 \pm 0.007 \right)\,  {\rm ps} \nonumber \\
%\tau(B^+) & = & \left(1.638 \pm 0.004 \right)\,  {\rm ps} \nonumber \\
%\tau(B_s^+) & = & \left(1.512 \pm 0.007 \right)\,  {\rm ps} \nonumber
%$\tau_{B^+}$ & $1.638 \pm 0.004$ \\
%$\tau_{B_{s}}$ & $1.512 \pm 0.007$ \\
%$\tau_{D^+}$ & $1.040 \pm 0.007$ \\
%$\tau_{D_s^+}$ & $0.500 \pm 0.007$\\
%\hline
%\end{tabular}
%\\
\end{center}
\caption{{\it Meson masses and lifetimes. The vector meson resonances $m_H^{\ast}$ are used in the extrapolation formula for $q^2$-dependence 
of the form factors (6.1). }}
\label{tab-6}
\end{table}
%%%%%%%%%%%%%%%%%%%%%%%%%%%%%%%%%%%%%%%%%%%%%%%%%%
%%%%%%%%%%%%%%%%%%%%%%%%%%%%%%%%%%%%%%%%%%%%%%%%%

%%%%%%%%%%%%%%%%%%%%%%%%%%%%%%%%%%%%%
%
%   REFERENCES 
%
%%%%%%%%%%%%%%%%%%%%%%%%%%%%%%%%

\end{document}